\documentclass[12pt]{article}
\usepackage{amsmath,amssymb,amsthm,amsxtra,overpic,bm,epsfig,ulem,multirow} 
\usepackage{color,cite}
\usepackage{epstopdf}
\usepackage{booktabs}
\usepackage{footnote}
\usepackage{rotating}
\usepackage{hyperref}
\usepackage{url}
\usepackage{cite}
\usepackage{tikz}
\usetikzlibrary{arrows,shapes,snakes,automata,backgrounds,petri}
\usetikzlibrary{decorations.pathmorphing}
\usetikzlibrary{decorations.markings}
\usepackage{lineno}

\usetikzlibrary{shapes.geometric,arrows.meta,positioning,calc,fit}

\def\thefootnote{\fnsymbol{footnote}}

\usepackage{array}

\begin{document}

\vspace{0.2cm}

\begin{center}
{\Large\bf Precision three-Dimensional Atmospheric Neutrino Flux Calculation Based on Honda Flux Model 
} 

\end{center}

\vspace{0.2cm}

\begin{center}
{\bf Jie Cheng~$^{a}$}~\footnote{Email: chengjie@ncepu.edu.cn},
{\bf Yu-Feng Li~$^{b,~c}$}~\footnote{Email: liyufeng@ihep.ac.cn},
{\bf Liang-Jian Wen~$^{b}$}~\footnote{Email: wenlj@ihep.ac.cn}\\
\vspace{0.2cm}
{$^a$School of Nuclear Science and Engineering, North China Electric Power University, Beijing 102206, China}\\
{$^b$Institute of High Energy Physics, Chinese Academy of Sciences, Beijing 100049, China}\\
{$^c$School of Physical Sciences, University of Chinese Academy of Sciences, Beijing 100049, China}
\end{center}

\vspace{1.5cm}

\begin{abstract}
We present a comprehensive three-dimensional atmospheric neutrino flux calculation based on the well-recognized simulation framework develeped by Honda and his collaborators, incorporating for the first time the muon propagation inside the Earth and its subsequent decay or nuclear capture. Other updates of essential input models include: the AMS02-based primary cosmic ray model, IGRF2020 geomagnetic field, and muon-recalibrated hadronic interaction model. The calculation covers seven detector sites across diverse geomagnetic environments, spanning 10~MeV to $10^4$~GeV. Significant site-dependent differences appear at $E_\nu < 10$~GeV, with $\nu_\mu$ flux at IceCube approximately twice that at JUNO below 1~GeV. Compared to HKKMS15, deviations of 2\%--10\% are attributed to the updated input models. Below 100~MeV, we present precise flux results, revealing that muon propagation contributes a globally significant component to the low-energy neutrino flux at all sites, with an approximately site-independent absolute increment. The hadronic uncertainty is re-estimated across the energy range using the updated hadronic interaction model, with significant reduction of the systematic error compared to previous calculations. These results provide essential inputs for neutrino oscillation and rare-event search experiments including JUNO, Super-Kamiokande/Hyper-Kamiokande, DUNE, KM3NeT/ORCA, and IceCube, as well as direct dark matter detection experiments facing the neutrino fog.
\end{abstract}


\def\thefootnote{\arabic{footnote}}
\setcounter{footnote}{0}

\newpage

\section{Introduction}\label{sec:intro}

Atmospheric neutrinos ($\nu_{\rm atm}$) serve as a crucial probe across multiple frontiers of neutrino physics. In the neutrino oscillation studies, they provide direct sensitivity to the atmospheric mixing parameters $\Delta m^2_{32}$ and $\theta_{23}$, as well as to the neutrino mass ordering through matter effects, primarily at energies $E_\nu \sim 1$--$10$~GeV~\cite{Super-Kamiokande:2017yvm,Super-Kamiokande:2023ahc,Super-Kamiokande:2015qek,IceCube:2014flw,ANTARES:2018rtf,MINOS:2014rjg,Aiello:2024}. In searches for the diffuse supernova neutrino background (DSNB), atmospheric neutrinos constitute the dominant irreducible background at $E_\nu \sim 10$--$30$~MeV, demanding precise flux predictions in this low-energy regime~\cite{Super-Kamiokande:2025sxh,KamLAND:2021gvi,Borexino:2019wln,JUNO:2022lpc}. In direct dark matter searches, coherent neutrino-nucleus scattering constitutes the irreducible ``neutrino fog''~\cite{OHare:2021utq} that fundamentally limits sensitivity to the WIMP dark matter candidate, while atmospheric neutrinos are the dominant background for indirect dark matter searches across the full energy range: at high energies (GeV--TeV) via neutrinos from WIMP annihilation in the Sun or Galactic Center at neutrino telescopes such as IceCube~\cite{IceCube:2016dgk}, and at low energies (tens of MeV to GeV) via monoenergetic neutrinos from light dark matter annihilation in the Galactic halo at liquid-scintillator and water-Cherenkov detectors such as JUNO~\cite{JUNO:2023vyz}, Borexino~\cite{Borexino:2019wln}, and Super-Kamiokande (SK)~\cite{Super-Kamiokande:2020sgt}. Furthermore, atmospheric neutrinos are the irreducible background for nucleon decay searches, limiting the sensitivity of current and next-generation experiments including SK~\cite{Super-Kamiokande:2020wjk}, JUNO~\cite{JUNO:2022qgr}, Hyper-Kamiokande~\cite{Hyper-Kamiokande:2018ofw}, and DUNE~\cite{DUNE:2020ypp}. Accurate calculations of the $\nu_{\rm atm}$ flux across the full energy spectrum are therefore essential for all these physics programs.

A three-dimensional (3D) atmospheric neutrino flux calculation resolves the differential flux as a function of energy $E_\nu$, arrival direction $(\theta_z, \phi)$, and time, using full spherical geometry, a 3D geomagnetic field, and a spatiotemporally resolved atmospheric density model. Crucially, the transverse momentum $p_T$ of secondary particles is preserved and Earth's curvature is fully accounted for, enabling the capture of azimuthal asymmetries at sub-GeV energies and large zenith angles that are averaged out in lower-dimensional approaches. Several mature frameworks realize this 3D scheme via Monte Carlo simulation: the HKKMS~\cite{Honda:2004yz,Honda:2006qj,Honda:2011nf,Honda:2015fha}, Bartol~\cite{Barr:2004br}, and FLUKA~\cite{Battistoni:2002ew} calculations. An alternative semi-analytical approach, MCEq~\cite{Fedynitch:2018cbl,Kozynets:2023tsv}, solves cascade equations along each propagation direction and has been extended to two-dimensional angular distributions; while it can incorporate 3D geomagnetic cutoffs, the cascade evolution remains one-dimensional.
These calculations have been widely used in neutrino oscillation experiments, including SK~\cite{Super-Kamiokande:2017yvm,Super-Kamiokande:2023ahc,Super-Kamiokande:2015qek}, IceCube~\cite{IceCube:2014flw}, ANTARES~\cite{ANTARES:2018rtf}, and MINOS~\cite{MINOS:2014rjg}. Their optimization has been focused primarily on the $E_\nu > 1$~GeV regime relevant for oscillation physics, where systematic uncertainties are well controlled---for example, the HKKMS framework achieves $\lesssim 10\%$ uncertainty in the $1$--$100$~GeV range through atmospheric muon ($\mu_{\rm atm}$) flux calibration~\cite{Honda:2006qj,Sanuki:2006yd}. More recently, an alternative approach using accelerator-data-driven tuning of hadronic interactions has achieved 7--9\% flux uncertainty below 1~GeV~\cite{Sato:2026avb}. However, for $E_\nu < 100$~MeV---the energy range critical for DSNB and dark matter searches---three significant issues remain unaddressed: (i)~the input models have not been updated to reflect latest experimental measurements and the best description of the input models; (ii)~the contribution from atmospheric muons that propagate inside the Earth, stop, and subsequently decay or are captured by nuclei has never been considered; and (iii)~systematic site-to-site flux differences across different geomagnetic environments have not been comprehensively compared.

Among these gaps, the most fundamental is the absence of muon propagation inside the Earth. Atmospheric muons produced in hadronic cascades possess sufficient energy to penetrate the ground and enter the Earth's crust or ocean, where they rapidly lose energy and eventually come to rest. The subsequent decay or nuclear capture of these stopped muons produces a supplementary flux of low-energy neutrinos~\cite{Guo:2018sno}---a qualitatively new source that was entirely absent in all previous 3D flux calculations (HKKMS11~\cite{Honda:2011nf}, HKKMS15~\cite{Honda:2015fha}, Bartol~\cite{Barr:2004br}, FLUKA~\cite{Battistoni:2002ew}). Crucially, the nuclear capture channel produces $\nu_\mu$ with an energy spectrum extending to $\sim 95$~MeV, well above the Michel endpoint at $m_\mu/2 \approx 52.8$~MeV, directly impacting the DSNB and dark matter search regions.

In this work, we present a comprehensive 3D $\nu_{\rm atm}$ flux calculation based on the HKKMS15 framework~\cite{Honda:2015fha}, with the following key improvements:

\begin{enumerate}

\item We incorporate for the first time the propagation of $\mu_{\rm atm}$ inside the Earth and their subsequent decay or nuclear capture, which produces an additional flux of low-energy neutrinos below $\sim 100$~MeV.

\item We update all essential input models to reflect recent experimental measurements: the AMS02-based primary cosmic ray model, IGRF2020~\cite{IGRF13} for the geomagnetic field, and the muon-recalibrated hadronic interaction model.

\item The calculation is extended to a global network of seven detector sites (JUNO~\cite{JUNO:2021vlw}, SK~\cite{Super-Kamiokande:2017yvm}, CJPL~\cite{Cheng:2017usi}, KM3NeT/ORCA~\cite{Aiello:2024}, IceCube~\cite{IceCube:2014flw}, DUNE~\cite{DUNE:2020ypp}, TRIDENT~\cite{Ye:2023dch}), enabling a systematic comparison of fluxes across different geomagnetic environments, from near-zero field at the South Pole to $\sim 40{,}000$~nT at low-latitude sites.

\item Precise $\nu_{\rm atm}$ flux results below 100~MeV are presented for these seven sites for the first time, covering an energy range from 10~MeV to $10^4$~GeV.

\item We update the hadronic uncertainty estimation using the muon-constrained method of Honda et al.~\cite{Honda:2019ymh}, achieving significant improvement across the energy range below 100~GeV.

\end{enumerate}

The remaining part of this paper is organized as follows. Sec.~\ref{sec:framework} describes the calculation framework, including the 3D scheme overview and the treatment of muon propagation inside the Earth. Sec.~\ref{sec:results} presents the flux results and discussion, divided into two energy regions: $E_\nu > 100$~MeV (multi-site comparison, zenith-angle dependence, azimuthal distribution, and comparison with HKKMS15) and $E_\nu < 100$~MeV (muon-propagation contribution). Sec.~\ref{sec:uncertainty} provides the updated flux model uncertainty estimation. Sec.~\ref{sec:future} discusses possible refinements and future directions. Sec.~\ref{sec:summary} summarizes the main results and conclusions.

\section{Atmospheric Neutrino Flux Calculations}\label{sec:framework}

Primary cosmic rays propagate through the geomagnetic field to reach the atmosphere, where they collide with atomic nuclei, triggering a cascade shower. This process produces numerous mesons, which subsequently decay into $\mu$ and $\nu_{\mu}$. The $\mu$ then decay further, generating $\nu_{e}$ and additional $\nu_{\mu}$. This series of phenomena, influenced by the geomagnetic field, is referred to as the $\nu_{\rm atm}$ and $\mu_{\rm atm}$ generation process. When simulating the generation processes, the procedure can be summarized in three main parts: First, consider the primary cosmic ray flux ($\Phi_{\mathrm{cr}}$); next, account for the influence of the geomagnetic field on charged particles ($R_{\mathrm{geo}}$); finally, evaluate the neutrino and muon yields ($Y_{\nu}$ and $Y_{\mu}$). This sequence can be described using the integral expression in Eq.~(\ref{Eq:eq1}).

\begin{equation}
    \Phi_{\nu} (\Phi_{\mu}) = \Phi_{\mathrm{cr}} \otimes R_{\mathrm{geo}} \otimes Y_{\nu} (Y_{\mu}) 
    \label{Eq:eq1}
\end{equation}

To reach the atmosphere and interact, the primary cosmic rays must first traverse the geomagnetic field. Thus, $R_{\mathrm{geo}}$ encompasses the filtering effect of the geomagnetic field, which is determined by the rigidity of cosmic ray particles, defined as $R_{\mathrm{cr}} \equiv P/(Ze)$, where $P$ is the momentum and $Z$ is the atomic number of the cosmic ray particle. The yields $Y_{\nu}$ and $Y_{\mu}$ depend on the hadronic interaction model, air profile, and meson-muon decay processes. Neutrinos primarily originate from the two-body decay modes of pions and kaons, as well as the subsequent muon decays. The decay chain from pions is given by $\pi^{+}(\pi^{-}) \rightarrow \mu^{+} (\mu^{-}) + \nu_{\mu}(\bar{\nu}_{\mu})$, followed by $\mu^{+} (\mu^{-}) \rightarrow e^{+} (e^{-}) + \nu_{e}(\bar{\nu}_{e}) + \bar{\nu}_{\mu}(\nu_{\mu})$. A similar chain occurs for charged kaons. Under conditions where all particles decay, we expect the following ratios: $(\nu_{\mu}+\bar{\nu}_{\mu})/(\nu_{e}+\bar{\nu}_{e}) \sim 2$, $\nu_{\mu}/\bar{\nu}_{\mu} \sim 1$ and $\nu_{e}/\bar{\nu}_{e} \sim \mu^{+}/\mu^{-}$. Additionally, the kinematic of $\pi$ and $\mu$ decay results in each neutrino in the decay chain carrying roughly equal energy on average.

\begin{figure}[!t]
\begin{center}
\includegraphics[width=0.9\textwidth]{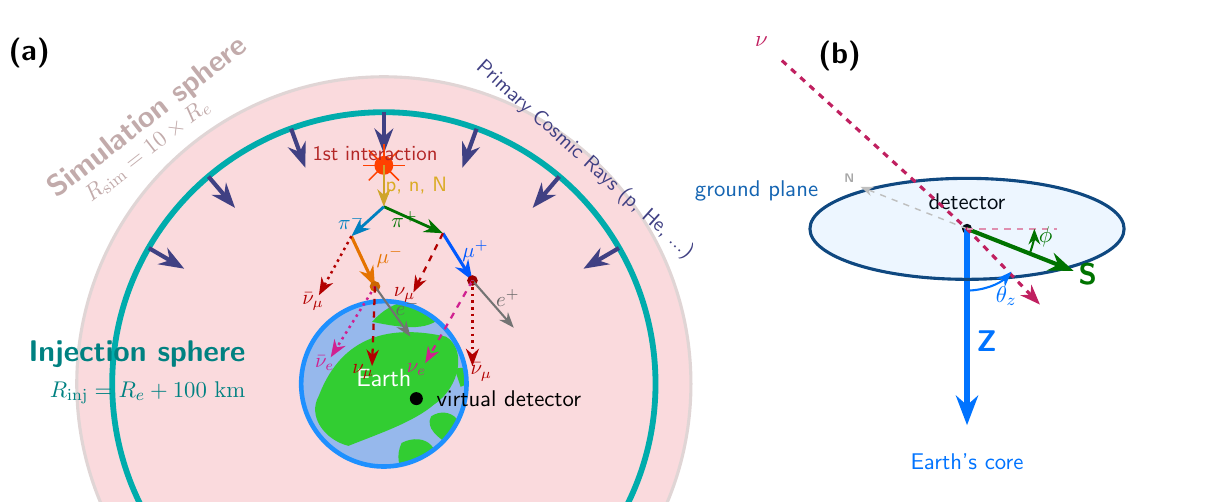}
\end{center}
\vspace{-0.8cm}
\caption{(a) Schematic illustration of the atmospheric neutrino flux calculation scheme (not to scale). (b) Definition of the arrival zenith angle $\theta_z$ and azimuth angle $\phi$ in the local detector coordinate system. $\theta_z$ is measured between the downward vertical (pointing toward the Earth's core) and the neutrino arrival direction; $\phi$ is measured counter-clockwise from the South direction in the horizontal plane.}
\label{fig:fig3d}
\end{figure}

The framework of the HKKMS flux calculation has established a systematic 3D scheme~\cite{Honda:2004yz,Honda:2006qj,Honda:2011nf,Honda:2015fha} for calculating $\nu_{\rm atm}$ flux, as shown in Fig.~\ref{fig:fig3d} (a).
In this scheme, the surface of the Earth is modeled as a sphere with a radius of $R_{e} = 6378.18$~km. To account for the geomagnetic barrier, two additional concentric spheres are defined as follows: the injection sphere ($R_{\mathrm{inj}} = R_{e} + 100$~km) and the simulation sphere ($R_{\mathrm{sim}} = 10 \times R_{e}$). 
Cosmic rays are sampled on the injection sphere and traced backwards in time to test the geomagnetic boundary condition. A particle is accepted if its trajectory reaches the simulation sphere ($R_{\mathrm{sim}}$) without re-contacting the injection sphere ($R_{\mathrm{inj}}$). The subsequent atmospheric cascade propagation is simulated in the region between the Earth's surface ($R_e$) and the simulation sphere. Additionally, the current framework tracks charged secondaries (primarily muons) that penetrate the Earth. As they lose energy and eventually stop within the Earth's material, their subsequent decay or capture produces a supplementary flux of low-energy neutrinos. This process, absent in previous HKKMS versions, is detailed in Appendix~\ref{app:muon}. 
Neutrino detectors are treated as infinitesimal points on the Earth's surface. In the 3D calculation, a finite-size ``virtual detector'' is defined for each target, covering a circular area of radius $\theta_{d}$ around the actual detector site. Neutrinos passing through this area are registered, and a ``virtual detector correction'' is applied to account for the finite size effects~\cite{Honda:2006qj,Honda:2011nf,Honda:2015fha}.

Additionally, determining neutrino directions is a critical aspect of the calculation. The definitions of arrival zenith angle and azimuth angle are shown in Fig.~\ref{fig:fig3d} (b). The arrival zenith angle is defined by measuring the angle between two directions: one pointing downward vertically (as a vector extending from Earth's core through the neutrino’s location in the observation area) and the other representing the direction from which the neutrino arrives. The azimuth angle is measured counter-clockwise from the south.


\begin{table}[!tb]
\centering
\caption{
Summary of the main features of HKKMS11, HKKMS15 and current calculations for atmospheric neutrino flux.}
\begin{tabular}{p{5cm}| c| c| c}
\hline \hline
Calculations &  HKKMS11  &  HKKMS15  &  This work \\
\hline \hline
Primary cosmic ray model & \multicolumn{2}{c|}{AMS01-based} & AMS02-based   \\
\hline
Geomagnetic field model & IGRF2005  & IGRF2010 & IGRF2020\\
\hline
Atmospheric model & US-standard '76 & \multicolumn{2}{c}{NRLMSISE-00} \\
\hline
\multirow{2}{*}{Hadronic interaction model} & \multicolumn{2}{c|}{JAM / DPMJET-III} & JAM / DPMJET-III \\
       & \multicolumn{2}{c|}{(muon-calibrated)} & (muon-recalibrated)\\
\hline
Inclusion of $\mu$ propagation inside the Earth & \multicolumn{2}{c|}{$\times$} & \checkmark \\

\hline

\hline \hline
\end{tabular}
\label{table:tab1}
\end{table}

Building upon the HKKMS15 calculation scheme~\cite{Honda:2015fha}, we have performed a comprehensive 3D $\nu_{\rm atm}$ flux calculation with several key improvements. First, the calculation is extended to a global network of seven detector sites: JUNO, SK, CJPL, KM3NeT/ORCA, IceCube, DUNE, and TRIDENT. Second, we incorporate the aforementioned muon propagation process inside the Earth, which significantly contributes to the neutrino flux below 100~MeV. Finally, we have updated the essential input models to reflect recent experimental measurements. Table~\ref{table:tab1} summarizes the key features of the HKKMS11, HKKMS15, and current calculations, highlighting these model variations. The specific models and data inputs for Eq.~(\ref{Eq:eq1}) used in this work are detailed in the following subsections.

The four input models used in Eq.~(\ref{Eq:eq1}) are updated compared to HKKMS15, as summarized in Table~\ref{table:tab1}. The AMS02-based primary cosmic ray model incorporates measurements from AMS02~\cite{AMS:2015tnn,AMS:2015azc}, BESS-polar~\cite{Abe:2015mga}, and PAMELA~\cite{PAMELA:2011mvy} at low energies, and JACEE~\cite{Christ:1998zz}, RUNJOB~\cite{RUNJOB:2005mtb}, and CREAM~\cite{Yoon:2017qjx} above 1~TeV, yielding higher fluxes below $\sim 40$~GeV and lower fluxes above compared to the AMS01-based model; the impact on the calculated $\nu_{\rm atm}$ flux is discussed in Sec.~\ref{subsec:comparison_hkkms15}. The geomagnetic field is described by IGRF2020~\cite{IGRF13}, whose site-dependent rigidity cutoff and muon bending corrections are essential for accurate flux predictions across the seven detector sites. The NRLMSISE-00 global atmospheric model~\cite{NR00} provides site- and time-dependent air density profiles, with the resulting systematic uncertainty estimated at $\sim 3\%$ for mid-latitude sites~\cite{Honda:2006qj}. For hadronic interactions, the JAM and modified DPMJET-III combination is employed with the muon-recalibrated tuning, which reduces the $\mu_{\rm atm}$ flux discrepancy to within 5\% over 1--100~GeV. Full details of each model are provided in Appendix~\ref{app:models}, including the primary cosmic ray spectra (Fig.~\ref{fig:crmodel}), the geomagnetic field map (Fig.~\ref{fig:geomagnetic_field}), and the hadronic calibration procedure.

A key innovation of this work is the inclusion of muon propagation inside the Earth. Atmospheric muons that penetrate the ground rapidly lose energy and eventually come to rest. Once stopped, a $\mu^{+}$ decays freely ($\mu^{+} \to e^{+} \nu_{e} \bar{\nu}_{\mu}$), while a $\mu^{-}$ is captured into a muonic atom where it either decays in orbit ($\mu^{-} \to e^{-} \bar{\nu}_{e} \nu_{\mu}$) or undergoes nuclear capture ($\mu^{-} + p \to \nu_{\mu} + n$). The branching between these channels is determined by element-dependent probabilities~\cite{Guo:2018sno}. Crucially, nuclear capture produces $\nu_{\mu}$ with an energy spectrum extending to $\sim$95~MeV, well above the Michel endpoint at $m_{\mu}/2 \approx 52.8$~MeV. The stopping medium (rock or water) is identified using the CRUST1.0 global crustal model~\cite{crust}. A comprehensive description of the decay spectra, capture probabilities, and the computational workflow is provided in Appendix~\ref{app:models}.
The limitations of the present framework and possible future refinements are discussed in Sec.~\ref{sec:future}.

\section{Results and Discussion}\label{sec:results}

The current calculation yields the full three-dimensional $\nu_{\rm atm}$ flux across seven detector sites (JUNO, SK, CJPL, KM3NeT/ORCA, IceCube, DUNE, and TRIDENT), spanning an energy range from 10~MeV to $10^4$~GeV. The resulting flux predictions are analyzed in two energy regimes, each motivated by distinct physics applications.
\textbf{Sec.~\ref{sec:flux_high_energy} ($E_\nu > 100$~MeV)} addresses the flux predictions relevant for neutrino oscillation studies and flux measurements. We present the all-direction-averaged fluxes and their site-dependent variations, examine the zenith-angle and azimuth-angle dependence, trace the contributions of primary cosmic rays of different energies, and compare our results with the HKKMS15 calculation to quantify the combined impact of updated input models.
\textbf{Sec.~\ref{subsec:elow100} ($E_\nu < 100$~MeV)} focuses on the low-energy regime critical for DSNB and dark matter searches. This region had not been precisely calculated in previous 3D flux frameworks. We first establish the baseline flux without muon propagation, then quantify the additional contribution from muon propagation inside the Earth, and analyze its global characteristics and implications for multi-site flux comparisons.

\subsection{Region of $E_\nu > 100$ MeV}~\label{sec:flux_high_energy}
\subsubsection{All-Direction Averaged Fluxes and Ratios}

\begin{figure}[!t]
\begin{center}
\includegraphics[width=0.9\textwidth]{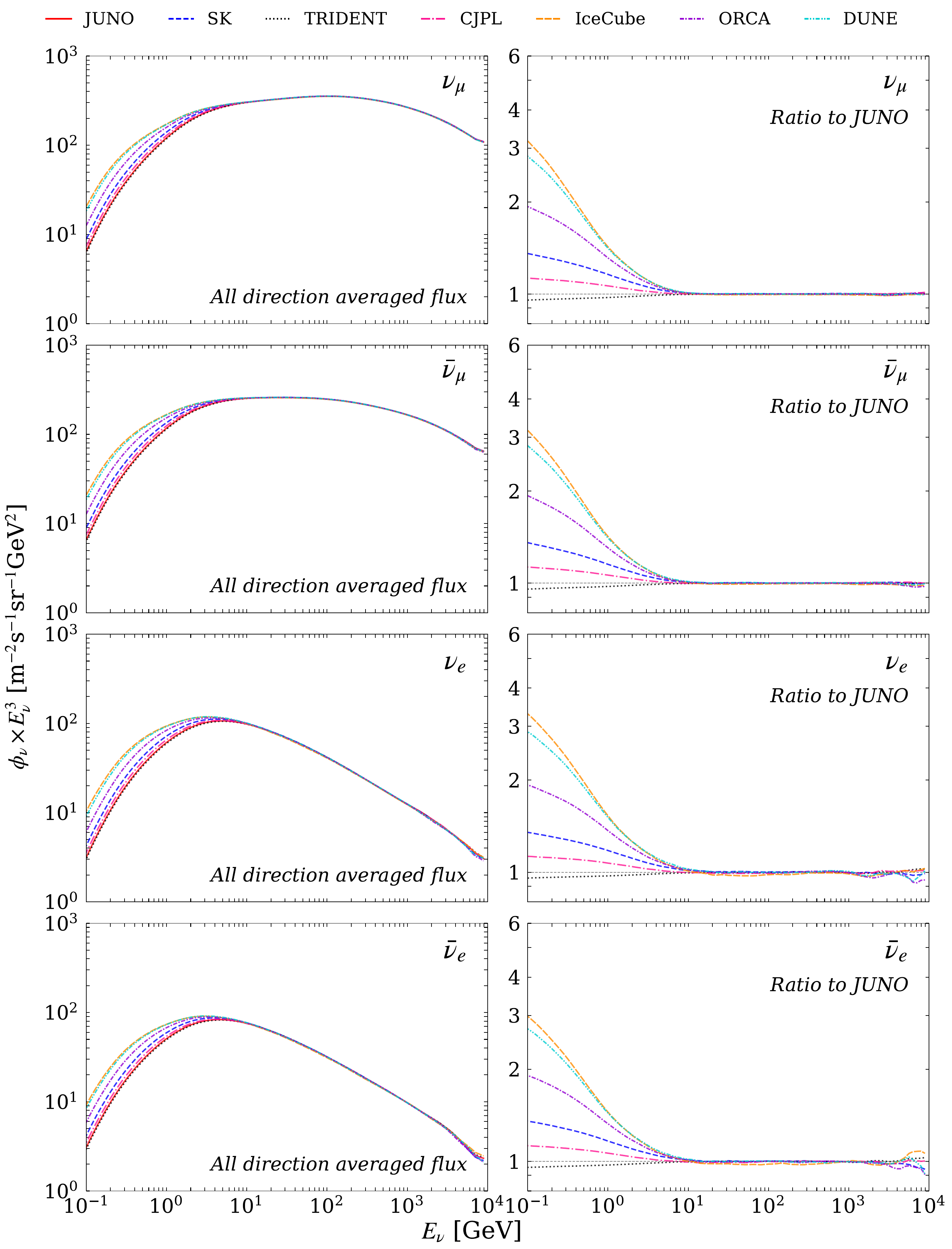}
\end{center}
\vspace{-0.8cm}
\caption{Yearly averaged, all-direction averaged atmospheric neutrino fluxes predicted for seven distinct detector sites: JUNO, SK, TRIDENT, IceCube, CJPL, KM3NeT/ORCA, and DUNE. Left panels: The differential flux scaled by the cube of the energy ($\phi_\nu \times E_\nu^3$) as a function of $E_\nu$ on a logarithmic scale from $10^{-1}$ to $10^4$ GeV. Right panels: Corresponding flux ratios with respect to the JUNO prediction.}
\label{fig:avgflux}
\end{figure}

\paragraph{Flux spectra.}
Figure~\ref{fig:avgflux} presents the yearly averaged, all-direction averaged $\nu_{\rm atm}$ fluxes predicted for seven distinct detector sites: JUNO, SK, TRIDENT, IceCube, CJPL, KM3NeT/ORCA, and DUNE. The left panels display the differential flux scaled by the cube of the energy ($\phi_\nu \times E_\nu^3$) as a function of $E_\nu$ on a logarithmic scale from $10^{-1}$ to $10^4$ GeV. The four rows correspond to the four neutrino flavors: $\nu_\mu$, $\bar{\nu}_\mu$, $\nu_e$, and $\bar{\nu}_e$. The flux spectra for all sites exhibit a similar shape, rising at low energies, peaking between 1 and 10 GeV, and falling off at higher energies. While the curves overlap significantly at high energies, distinct variations are visible at lower energies ($E_\nu < 10$ GeV), reflecting differences in geographical location and geomagnetic cutoff effects. The right panels illustrate the ratio of the fluxes at the six other sites relative to the JUNO prediction. These panels quantify the site-dependent variations. At high energies ($E_\nu > 10$ GeV), the ratios for all sites converge to unity, indicating that the $\nu_{\rm atm}$ flux becomes independent of location. However, at lower energies, significant deviations occur. For example, in the $\nu_\mu$ channel, the ratios for IceCube and DUNE rise sharply towards 2.0 at energies below 1 GeV, whereas SK and CJPL show more moderate enhancements relative to JUNO.

\begin{figure}[!t]
\begin{center}
\includegraphics[width=0.9\textwidth]{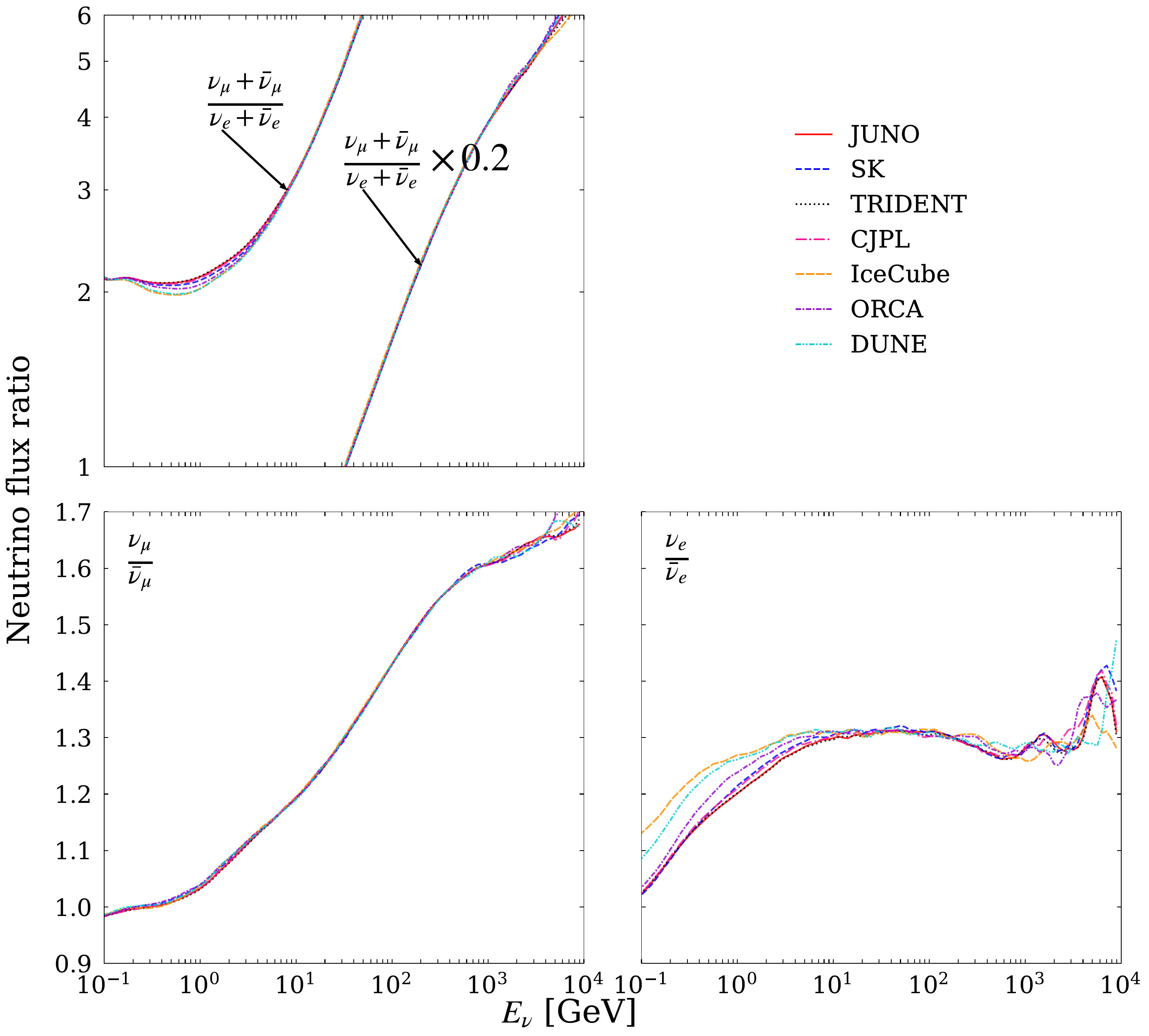}
\end{center}
\vspace{-0.8cm}
\caption{Neutrino flavor ratios for seven detector sites (JUNO, SK, CJPL, TRIDENT, IceCube, KM3NeT/ORCA, and DUNE), calculated using the all-directional, one-year-averaged atmospheric neutrino flux.}
\label{fig:nuratio}
\end{figure}

\paragraph{Flavor and charge ratios.}
Figure~\ref{fig:nuratio} presents the neutrino flavor and neutrino-antineutrino flux ratios for seven detector sites (JUNO, SK, CJPL, TRIDENT, IceCube, KM3NeT/ORCA, and DUNE), calculated using the all-directional, one-year-averaged $\nu_{\rm atm}$ flux.
The \textbf{top panel} displays both the ratio of the total muon-type neutrino flux to the total electron-type neutrino flux, $(\nu_\mu + \bar{\nu}_\mu)/(\nu_e + \bar{\nu}_e)$, and the same ratio scaled by a factor of 0.2. Assuming that all mesons and muons decay, this ratio is expected to be $\sim 2$ at low energies. The panel confirms this, with curves for all sites starting near 2 at $E_\nu \sim 0.1\ \mathrm{GeV}$ (consistent with dominant production via pion decay) and rising steeply at high energies. The indistinguishability of the curves across all sites indicates that the flavor composition is independent of geographical location.
The \textbf{bottom panels} illustrate the charge asymmetry ratios.
\begin{itemize}
    \item The bottom-left panel shows the ratio $\nu_\mu/\bar{\nu}_\mu$. Consistent with theoretical expectations for full decay, this ratio approaches $\sim 1$ at low energies and rises monotonically to approximately 1.7 at $10^4\ \mathrm{GeV}$.
    \item The bottom-right panel shows the ratio $\nu_e/\bar{\nu}_e$. This ratio directly reflects the $\pi^+/\pi^-$ ratio of the parent pions. A significant divergence among sites is observed below $10\ \mathrm{GeV}$, driven by variations in the local geomagnetic cutoff rigidity:
    \begin{description}
        \item[Low Cutoff Rigidity Regions:] Primary cosmic ray protons dominate, producing an excess of $\pi^+/\pi^-$. This results in a higher $\nu_e/\bar{\nu}_e$ ratio even at low energies (seen in the curves for IceCube and DUNE).
        \item [High Cutoff Rigidity Regions:] The contribution of secondary cosmic rays (particularly secondary neutrons) becomes significant. These interactions dilute the $\pi^+$ excess, thereby lowering the $\nu_e/\bar{\nu}_e$ ratio (seen in the curves for JUNO and SK).
    \end{description}
    At higher energies ($E_\nu > 10\ \mathrm{GeV}$), the curves for all sites converge.
\end{itemize}

\subsubsection{Zenith-Angle Dependence}

In the following, three representative zenith angle bins are adopted: downward-going ($1 > \cos\theta_z > 0.9$), horizontal ($0.1 > \cos\theta_z > 0.0$), and upward-going ($-0.9 > \cos\theta_z > -1$). These definitions are used throughout the following discussion unless otherwise noted.

\begin{figure}[!t]
\begin{center}
\includegraphics[width=1.0\textwidth]{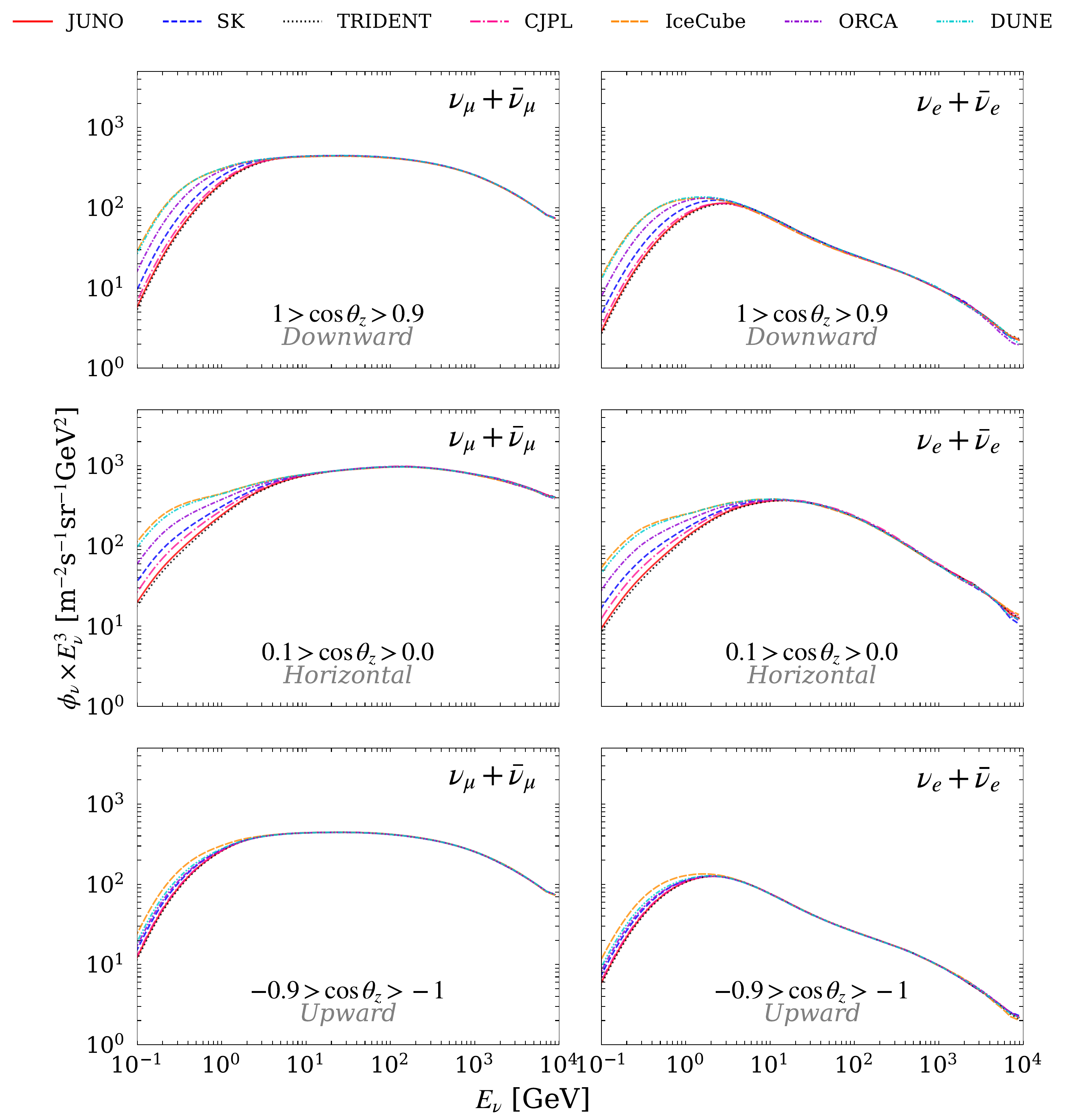}
\end{center}
\vspace{-0.8cm}
\caption{Yearly averaged atmospheric neutrino fluxes predicted for seven distinct detector sites (JUNO, SK, TRIDENT, IceCube, CJPL, KM3NeT/ORCA, and DUNE), binned in zenith angle and averaged over all azimuthal angles. Each panel shows the differential flux scaled by the cube of the energy ($\phi_\nu \times E_\nu^3$) as a function of $E_\nu$ on a logarithmic scale from $10^{-1}$ to $10^4$~GeV. The three rows correspond to three zenith angle bins: downward-going ($1 > \cos\theta_z > 0.9$), horizontal ($0.1 > \cos\theta_z > 0.0$), and upward-going ($-0.9 > \cos\theta_z > -1$). The two columns correspond to $\nu_\mu+\bar{\nu}_\mu$ and $\nu_e+\bar{\nu}_e$ (combined particle and antiparticle).}
\label{fig:flux_zenith}
\end{figure}

\begin{figure}[!t]
\begin{center}
\includegraphics[width=1.0\textwidth]{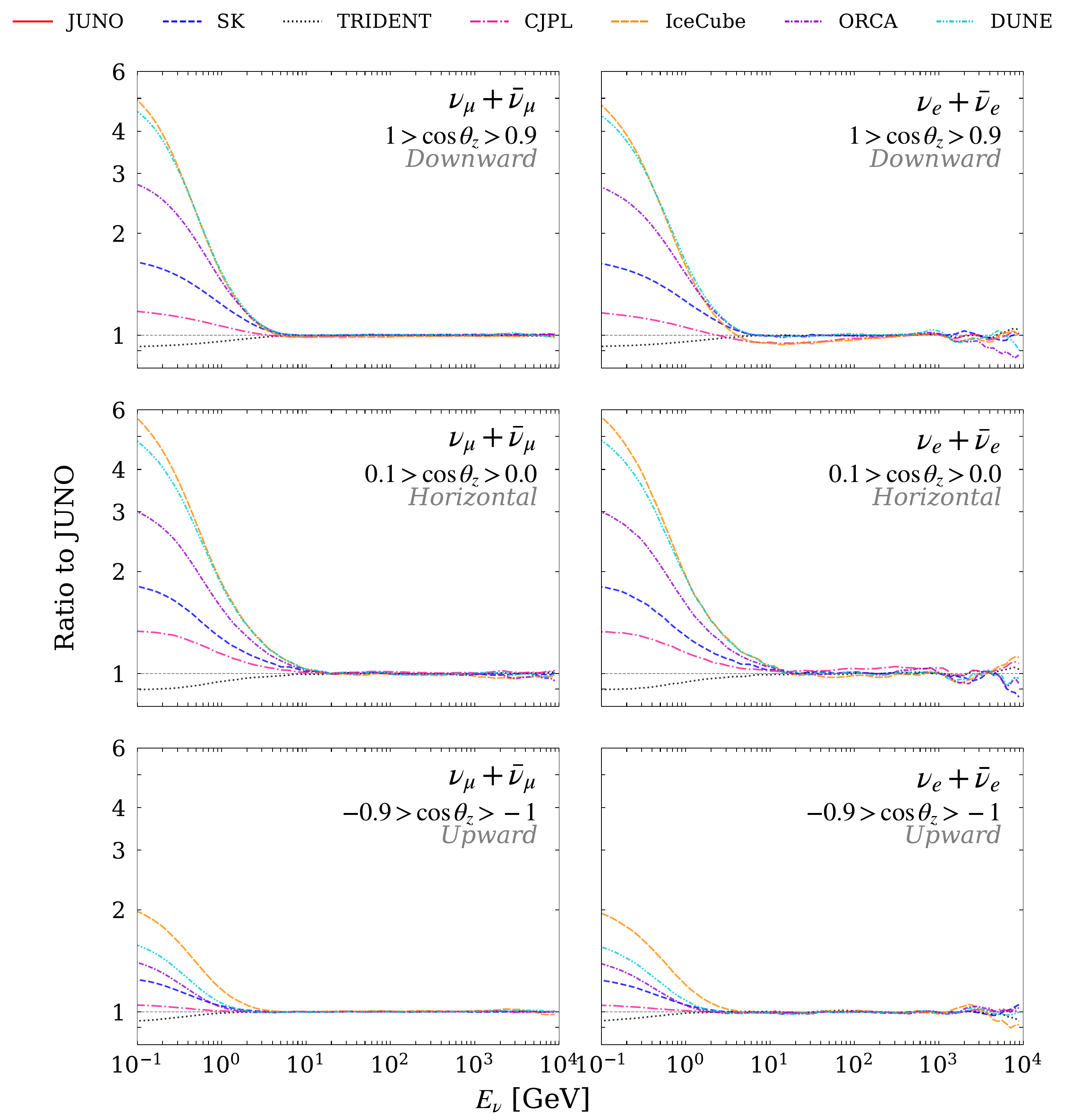}
\end{center}
\vspace{-0.8cm}
\caption{Ratios of yearly averaged atmospheric neutrino fluxes at seven distinct detector sites (SK, TRIDENT, IceCube, CJPL, KM3NeT/ORCA, and DUNE) relative to JUNO, binned in zenith angle and averaged over all azimuthal angles. Each panel shows the flux ratio as a function of $E_\nu$ on a logarithmic scale from $10^{-1}$ to $10^4$~GeV. The three rows correspond to three zenith angle bins: downward-going ($1 > \cos\theta_z > 0.9$), horizontal ($0.1 > \cos\theta_z > 0.0$), and upward-going ($-0.9 > \cos\theta_z > -1$). The two columns correspond to $\nu_\mu+\bar{\nu}_\mu$ and $\nu_e+\bar{\nu}_e$ (combined particle and antiparticle).}
\label{fig:ratio_zenith}
\end{figure}

\begin{figure}[!t]
\begin{center}
\includegraphics[width=1.0\textwidth]{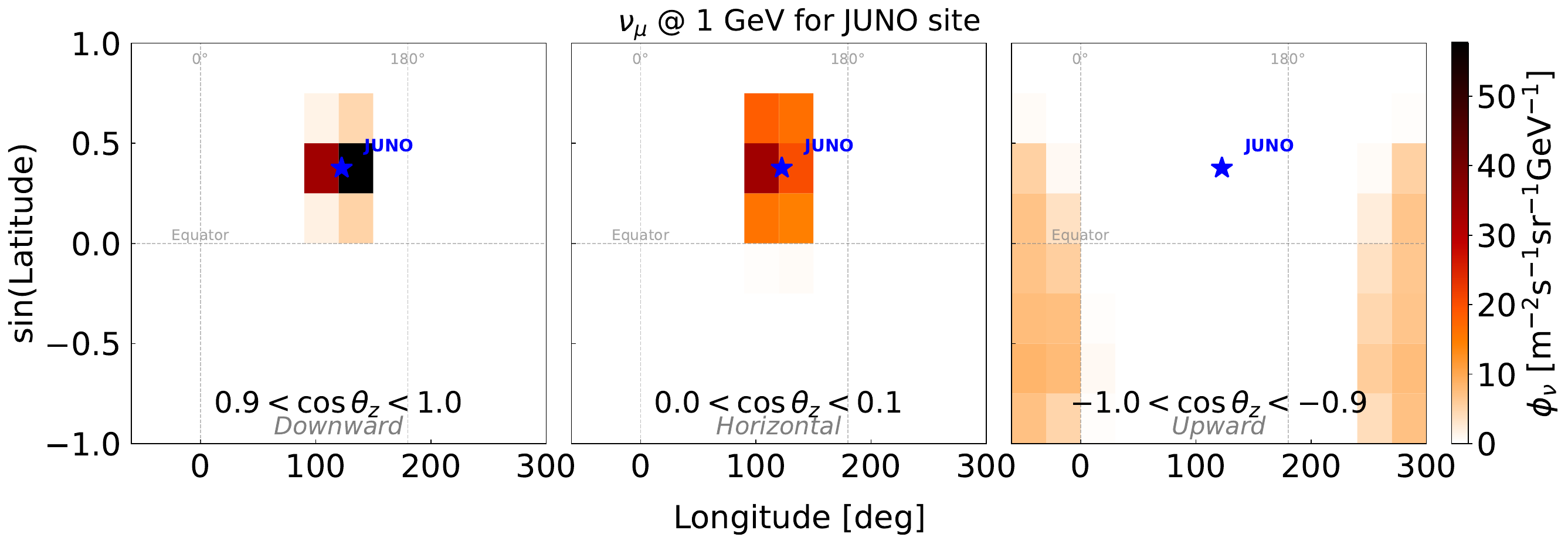}
\end{center}
\vspace{-0.8cm}
\caption{Flux-weighted production locations of 1 GeV atmospheric $\nu_\mu$ for the JUNO site, averaged over all azimuth angles, shown in geographic coordinates for three zenith angle bins: downward-going ($0.9 < \cos\theta_z < 1.0$), horizontal ($0.0 < \cos\theta_z < 0.1$), and upward-going ($-1.0 < \cos\theta_z < -0.9$). The star marker indicates the geographic position of the JUNO site.}
\label{fig:prod_loc1}
\end{figure}

\paragraph{Zenith-dependent flux variations across sites.}
Figures~\ref{fig:flux_zenith}, \ref{fig:ratio_zenith}, and \ref{fig:prod_loc1} together illustrate the zenith-angle dependence of $\nu_{\rm atm}$ flux predictions for seven distinct detector sites (JUNO, SK, TRIDENT, IceCube, CJPL, KM3NeT/ORCA, and DUNE) and the underlying production geometry. Figure~\ref{fig:flux_zenith} shows the yearly averaged differential flux scaled by the cube of the energy ($\phi_\nu \times E_\nu^3$) as a function of neutrino energy from $10^{-1}$ to $10^4$ GeV, binned in zenith angle and averaged over azimuth. Figure~\ref{fig:ratio_zenith} presents the corresponding flux ratios relative to JUNO. Figure~\ref{fig:prod_loc1} maps the flux-weighted production locations at 1 GeV for the JUNO site only, for three representative zenith angle bins. The star marker in each panel indicates the geographic position of the JUNO detector, providing a reference for the production regions. Figs.~\ref{fig:flux_zenith} and \ref{fig:ratio_zenith} are organized as 3 (zenith bins) $\times$ 2 ($\nu_\mu+\bar{\nu}_\mu$, $\nu_e+\bar{\nu}_e$) panels, while Fig.~\ref{fig:prod_loc1} shows only $\nu_\mu$.

The production location map for JUNO (Fig.~\ref{fig:prod_loc1}) provides essential context for understanding the site-dependent flux variations shown in Figs.~\ref{fig:flux_zenith} and \ref{fig:ratio_zenith}. Although Fig.~\ref{fig:prod_loc1} displays results for JUNO only, the production geometry is qualitatively similar for other sites, with the specific production region shifting according to each detector's geographic coordinates. For downward-going neutrinos, the production region is concentrated near the detector site, making the flux directly sensitive to the local geomagnetic field. Horizontal neutrinos also originate predominantly from regions near the detector, but their elongated atmospheric path makes them particularly sensitive to the local geomagnetic field. Upward-going neutrinos originate from the opposite side of the Earth, with production regions centered at the antipodal longitude relative to the detector. 
The different flux scales across the three geographic panels---highest for downward-going, intermediate for horizontal, and lowest for upward-going neutrinos---reflect the geometric dilution 
effect: downward-going neutrinos originate from a compact region directly above the detector, concentrating the flux into a small solid angle, while upward-going neutrinos are produced over a broad area on the opposite side of the Earth, with the flux diluted by the large propagation distance and wide spatial distribution. Notably, JUNO's mid-latitude location results in one of the lowest $\nu_{\rm atm}$ fluxes among the compared sites (surpassed only by TRIDENT), which is advantageous for reducing $\nu_{\rm atm}$ background in reactor antineutrino and DSNB analyses.

These production geometries directly explain the patterns observed in the flux ratios (Fig.~\ref{fig:ratio_zenith}). 
At low energies ($E_\nu \lesssim 5$ GeV), the flux ratio deviations are most pronounced in the horizontal bin and smallest in the upward-going bin. This ordering reflects how the production geometry interacts with the local geomagnetic field. Downward-going neutrinos are produced in a compact region directly above the detector and are sensitive to the local geomagnetic cutoff, leading to significant site-to-site differences. Horizontal neutrinos also originate predominantly from regions near the detector, but their elongated atmospheric path makes them even more susceptible to the local geomagnetic field, yielding the largest inter-site variation. Upward-going neutrinos, by contrast, are produced over a broad area on the opposite side of the Earth, where the geomagnetic field is effectively averaged over a wide range of latitudes, resulting in the smallest site-to-site differences. 

In the intermediate energy range ($\sim 5$--100 GeV), a distinct flavor-dependent suppression appears for $\nu_e$ and $\bar{\nu}_e$ in the down-going direction, with ratios for IceCube and CJPL dropping to $\sim 0.9$. Unlike $\nu_\mu$ and $\bar{\nu}_\mu$, which are produced directly in pion and kaon decays ($\pi/K \to \mu \nu_\mu$), electron neutrinos arise exclusively from secondary muon decay ($\mu \to e \nu_e \bar{\nu}_\mu$). At these energies, a substantial fraction of muons reach the ground before decaying, suppressing the $\nu_e$ and $\bar{\nu}_e$ flux. This effect is amplified at IceCube due to reduced air density from extreme cold, and at CJPL due to the shorter atmospheric overburden at 1580 m altitude. Both conditions increase the probability that muons survive to the ground rather than decaying in flight.

For upward-going neutrinos, the production regions are distant from the detector, the atmospheric path is long enough to ensure nearly complete muon decay, and the geomagnetic conditions are effectively averaged over the global field. As a result, the fluxes are nearly identical across all sites at all energies. In fact, above $\sim 100$ GeV, the fluxes converge to unity across all sites for all zenith angle bins and flavors, confirming the expected global symmetry of high-energy $\nu_{\rm atm}$ production.

\begin{figure}[!t]
\begin{center}
\includegraphics[width=0.9\textwidth]{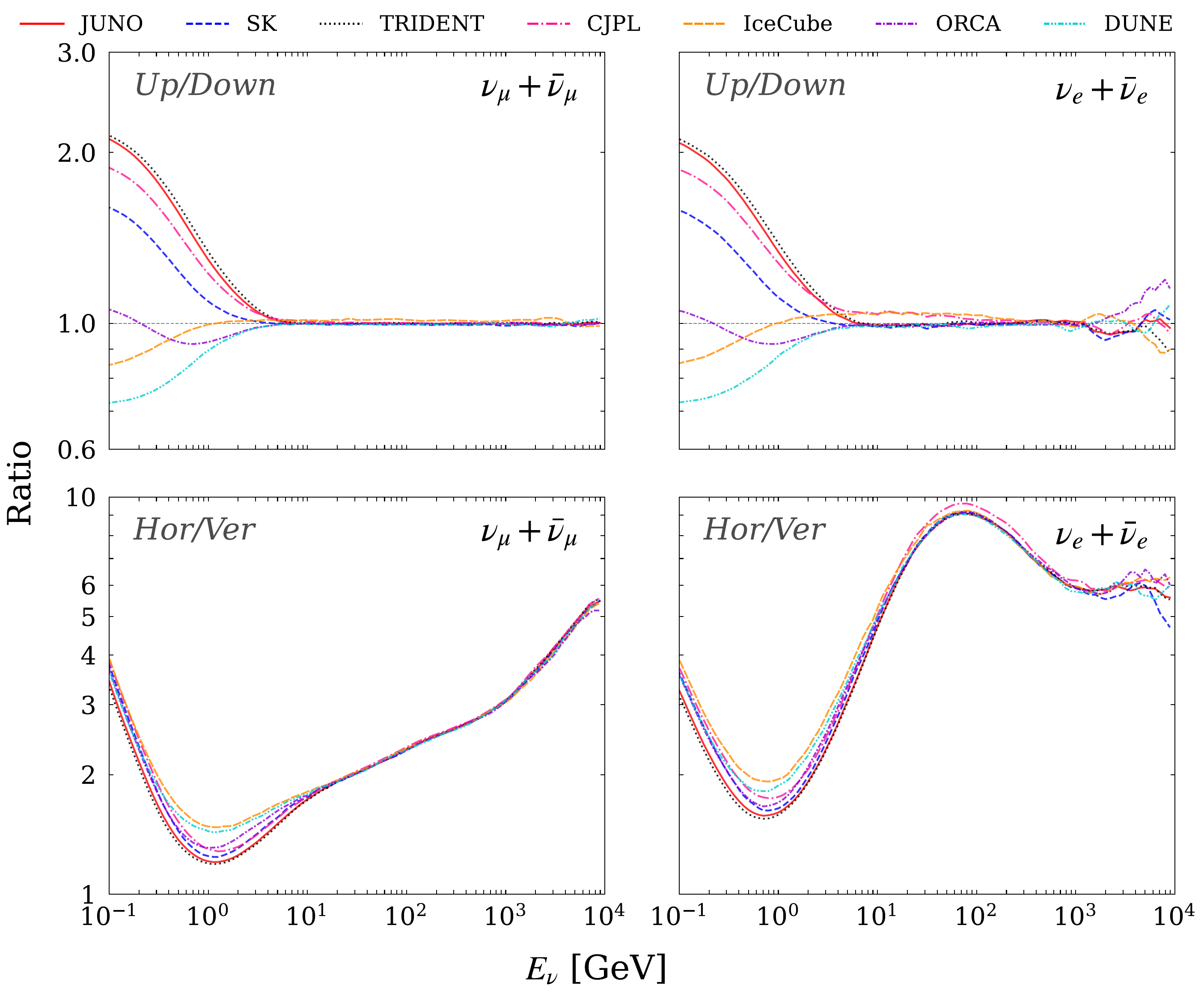}
\end{center}
\vspace{-0.8cm}
\caption{Zenith-angle flux ratios of atmospheric neutrinos for seven detector sites. The top row displays the Up-to-Down ratio ($-0.9 > \cos\theta_z > -1$ vs. $1 > \cos\theta_z > 0.9$), while the bottom row shows the Horizontal-to-Vertical ratio ($0.1 > \cos\theta_z > 0$ vs. $1 > \cos\theta_z > 0.9$).}
\label{fig:up2downandhor2ver}
\end{figure}

\paragraph{Up-to-Down and Horizontal-to-Vertical ratios.}
Figure~\ref{fig:up2downandhor2ver} presents the zenith-angle flux ratios for seven detector sites, quantifying the anisotropy of $\nu_{\rm atm}$ fluxes. The top row displays the Up-to-Down ratio (upward-going vs.\ downward-going), while the bottom row shows the Horizontal-to-Vertical ratio (horizontal vs.\ downward-going). All subplots display the sum of neutrino and antineutrino fluxes ($\nu + \bar{\nu}$). All panels use a logarithmic energy scale spanning from $0.1$ to $10^4$ GeV.

\medskip

\noindent\textbf{Up-to-Down Ratio (top panels of Fig.~\ref{fig:up2downandhor2ver}).}
In the low-energy regime ($E_\nu \lesssim 10$ GeV), the Up/Down ratios exhibit strong site-dependent deviations from unity, driven by geomagnetic shielding. As illustrated in Fig.~\ref{fig:prod_loc1}, downward-going neutrinos originate directly above the detector and sample the local geomagnetic cutoff rigidity, whereas upward-going neutrinos are produced at the antipodal region, sampling a globally averaged field. This geometric separation produces a wide spread in Up/Down ratios at $0.1$ GeV across the seven sites. For $\nu_\mu$ (top-left), values range from $\sim 2.2$ (TRIDENT) down to $\sim 0.7$ (DUNE), spanning a factor of $\sim 3$. For $\nu_e$ (top-right), the range is similar. The ranking reflects both geomagnetic latitude and the non-dipolar structure of Earth's magnetic field. TRIDENT (South China Sea, $\sim 15^\circ$N) exhibits the highest Up/Down ratio because its downward direction samples a very high local cutoff ($\sim 12$--$14$ GV) that strongly suppresses low-energy cosmic rays, while its upward direction samples the antipodal South Pacific ($\sim 15^\circ$S), where the effective cutoff is lower due to the offset of the geomagnetic dipole and the South Atlantic Anomaly. JUNO ($\sim 23^\circ$N) follows closely. Three northern-hemisphere sites---IceCube (Antarctica), KM3NeT/ORCA (Mediterranean, $\sim 43^\circ$N), and DUNE (Fermilab, $\sim 42^\circ$N)---have Up/Down ratios $< 1$, meaning their downward flux exceeds their upward flux. This reversal occurs because their local (downward) cutoff is moderate or low, while their antipodal regions lie in zones of higher effective cutoff: for IceCube, the antipode samples the Arctic region where the geomagnetic pole offset creates a non-zero cutoff; for KM3NeT/ORCA and DUNE, their antipodes fall in the South Pacific/South Atlantic where the IGRF field is stronger than the local northern-hemisphere direction. SK ($\sim 36^\circ$N) sits above unity ($\sim 1.2$) but below CJPL ($\sim 1.6$), because SK's antipode in South America lies closer to the South Atlantic Anomaly (reduced cutoff), whereas CJPL's antipode in Argentina samples a region of more typical southern-hemisphere cutoff.

A notable feature in the $\nu_e$ Up/Down ratio (top-right panel) is the sustained deviation above unity for CJPL and IceCube in the 10--100 GeV range. This arises from the lower atmospheric column density at high-altitude (CJPL) or polar-cold (IceCube) sites. The reduced density increases the probability of muons reaching the ground before decaying, which suppresses the $\nu_e$ flux in the vertical direction (shortest atmospheric path) more than in the upward direction (longest path), yielding an Up/Down ratio $> 1$.

Above $\sim 10$ GeV, all Up/Down ratios converge to unity for both flavors, confirming that geomagnetic and column-density effects become negligible. The site-specific curves collapse onto a single trajectory, indicating that the angular distribution of $\nu_{\rm atm}$ production becomes effectively isotropic in this regime.

\medskip

\noindent\textbf{Horizontal-to-Vertical Ratio (bottom panels of Fig.~\ref{fig:up2downandhor2ver}).}
The Hor/Ver ratio exhibits a richer energy evolution, with distinct behaviors across four regimes:
\begin{itemize}
\item Sub-GeV regime ($E_\nu \lesssim 1$ GeV). The Hor/Ver ratio decreases from its low-energy baseline toward unity. In this regime, the muon decay length ($L_{\text{decay}} = \gamma c \tau_\mu$) is short ($\sim$ hundreds of meters), so muons produced along the extended horizontal path ($\sim 40$ km) decay before reaching the detector, eliminating the horizontal advantage. The vertical direction, with its shorter path and higher cutoff rigidity at mid-latitude sites, yields a comparable flux, driving the ratio close to 1. At $0.1$ GeV, all seven sites exhibit Hor/Ver ratios in the range of $\sim 3$--$4$ for both flavors. The site-dependent spread is modest ($\sim 1$ unit), with the curves tightly clustered at this lowest energy. As energy increases, the ratio drops rapidly for all sites, reaching a minimum of $\sim 1.2$--$1.4$ near $1$ GeV before rising again into the horizontal-enhancement regime.

\item 1--100 GeV: horizontal enhancement. As energy rises, muon decay lengths grow to kilometers, and the elongated horizontal atmospheric path becomes an advantage: muons have a much higher probability of decaying in flight ($\mu \to e \nu_e \bar{\nu}_\mu$) along horizontal trajectories than along vertical ones. The Hor/Ver ratio climbs steadily, peaking at $\sim 7$ for $\nu_\mu$ and $\sim 10$ for $\nu_e$. Site-specific curves collapse onto a single trajectory above a few GeV, as the geomagnetic shielding effect fades and muon-decay kinematics dominate uniformly across all latitudes. The higher peak for $\nu_e$ reflects its exclusive origin from muon decay, whereas $\nu_\mu$ receives partial contributions from direct pion/kaon decays that are less sensitive to the path-length effect.

\item $\sim 100$ GeV: muon energy-loss turnover. The $\nu_e$ Hor/Ver ratio reaches a maximum near 100 GeV and then declines to a minimum around 1 TeV. This dip marks the onset of significant muon radiative energy loss (bremsstrahlung, pair production, photonuclear interactions). Horizontal muons traverse more atmospheric material before decaying, suffering greater energy depletion, which suppresses the horizontal $\nu_e$ flux. The $\nu_\mu$ Hor/Ver ratio, by contrast, continues to rise through this regime because $\nu_\mu$ is increasingly fed by direct kaon decay ($K \to \mu \nu_\mu$, critical energy $\epsilon_K \approx 850$ GeV), which bypasses the muon-decay bottleneck entirely.

\item $> 1$ TeV: prompt component and kaon dominance. The $\nu_\mu$ Hor/Ver ratio steepens its ascent above 1 TeV, reflecting the full dominance of kaon decays and the growing contribution of the prompt neutrino component from charm-hadron decays ($D \to \mu \nu_\mu X$). The $\nu_e$ Hor/Ver ratio, after reaching its minimum at $\sim 1$ TeV, begins to rise again above 10 TeV. This recovery is driven by prompt $\nu_e$ from charm decays ($D \to e \nu_e X$), which are produced essentially at the interaction point (charm lifetime $c\tau \sim 10^{-4}$ m) and are insensitive to atmospheric depth. Although the prompt component is intrinsically isotropic, it appears as an increase in the Hor/Ver ratio because the conventional $\nu_e$ component has been suppressed more strongly in the horizontal direction by muon energy loss. At $10$ TeV, both ratios reach $\sim 5$--$6$, with $\nu_\mu$ exhibiting a steeper rise than $\nu_e$ in the highest-energy bin.
\end{itemize}

\subsubsection{Primary Cosmic Ray Contributions to the Atmospheric Neutrino Flux}

\begin{figure}[!t]
\begin{center}
\includegraphics[width=0.9\textwidth]{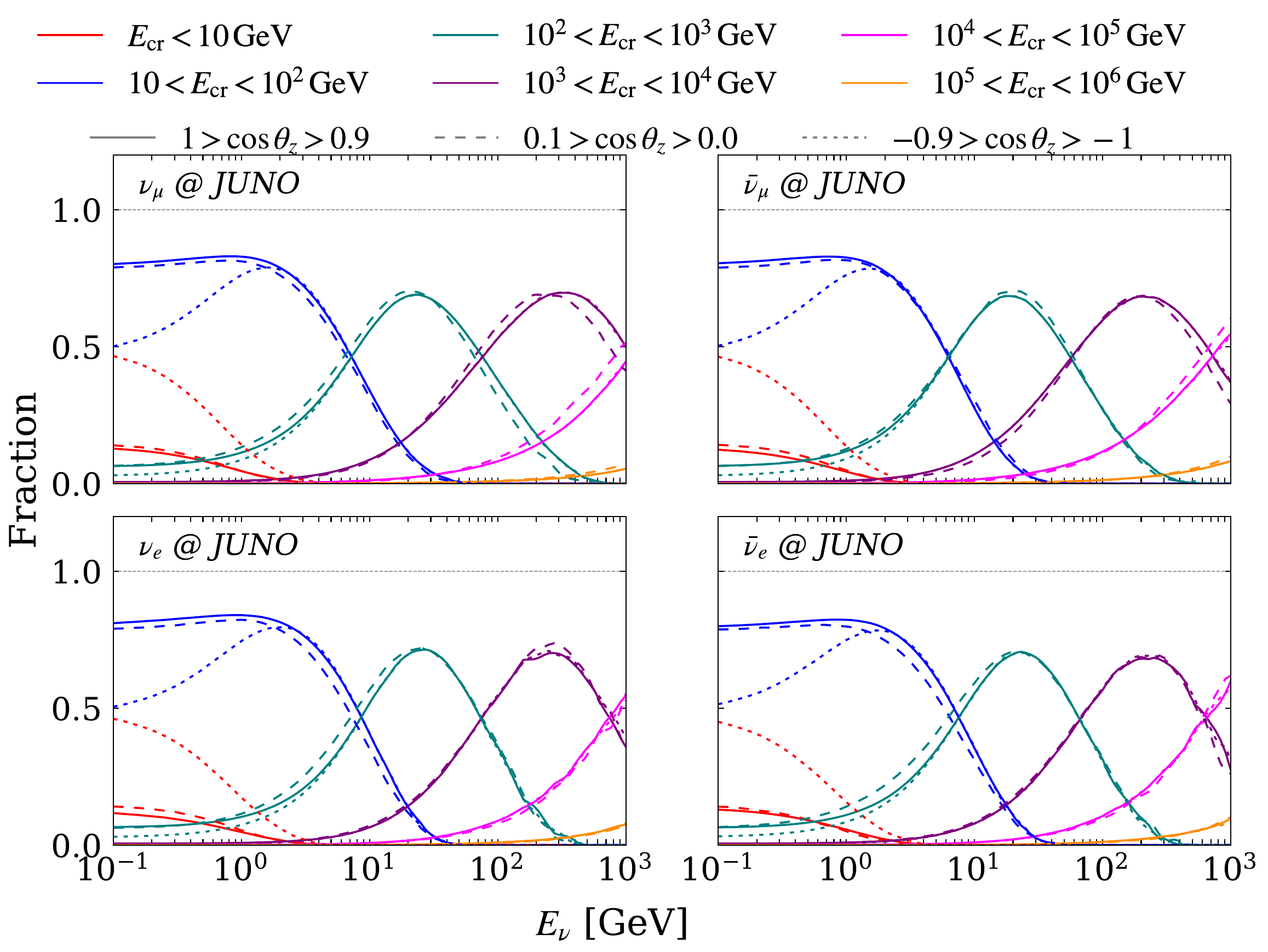}
\end{center}
\vspace{-0.8cm}
\caption{Fractional contribution of primary cosmic ray energy bins to the atmospheric neutrino flux at JUNO, for each flavor ($\nu_{\mu}$, $\bar{\nu}_{\mu}$, $\nu_{e}$, $\bar{\nu}_{e}$). Primary CR energy bins are color-coded from red ($E_{\rm cr} < 10$ GeV) to orange ($10^{5} < E_{\rm cr} < 10^{6}$ GeV). Line styles denote zenith angle ranges: solid ($1 > \cos\theta_{z} > 0.9$, down-going), dashed ($0.1 > \cos\theta_{z} > 0.0$, near-horizontal), and dotted ($-0.9 > \cos\theta_{z} > -1$, up-going).}
\label{fig:flux_ratio_by_Ecr}
\end{figure}

Having established the energy- and angle-dependent flux structure, we now examine how primary cosmic rays of different energies populate the $\nu_{\rm atm}$ spectrum. Figure~\ref{fig:flux_ratio_by_Ecr} shows the fractional contribution of different primary cosmic ray energy bins to the total $\nu_{\rm atm}$ flux at the JUNO site. Since event-level correspondence between primary cosmic rays and secondary neutrinos was not preserved in our simulation, we decompose the flux by the primary cosmic ray energy bin assigned during generation. This provides an approximate but useful picture of how primary cosmic rays of different energies populate the $\nu_{\rm atm}$ spectrum.

The four panels display the results separately for $\nu_\mu$, $\bar{\nu}_\mu$, $\nu_e$, and $\bar{\nu}_e$. Each curve represents the fraction of the total $\nu_{\rm atm}$ flux at a given neutrino energy $E_\nu$ originating from a specific primary cosmic ray energy bin $E_{\rm cr}$, such that the curves sum to unity at each $E_\nu$ for a given flavor and zenith angle. Here, $E_{\rm cr}$ denotes the kinetic energy of the primary cosmic ray.

A clear correlation between $E_{\rm cr}$ and $E_\nu$ is evident: lower-energy primaries predominantly produce lower-energy neutrinos, with the peak of each contribution shifting to higher $E_\nu$ as $E_{\rm cr}$ increases. Neutrinos in the sub-GeV range are produced almost exclusively by cosmic rays below 100 GeV, while those above 100 GeV require primaries above several TeV. In the intermediate region ($\sim$1--500 GeV), two or three adjacent $E_{\rm cr}$ bins contribute simultaneously, reflecting the broad energy spectrum of secondaries from hadronic cascades and muon decays. This overlap underscores the need for full energy convolution when mapping between primary cosmic ray and $\nu_{\rm atm}$ spectra.

The zenith-angle dependence reveals the influence of geomagnetic effects. At low neutrino energies, upward-going neutrinos (dotted curves) receive a larger fraction of their flux from the lowest $E_{\rm cr}$ bins compared to downward-going neutrinos (solid curves). This arises because low-energy cosmic rays are more readily deflected by the geomagnetic field, enhancing the flux of nearly horizontal and upward-arriving particles at mid-latitude sites. At higher energies, the three zenith-angle curves converge as cosmic rays become increasingly isotropic and the geomagnetic cutoff becomes negligible.

The four flavor panels exhibit nearly identical shapes, reflecting the common origin of $\nu_{\rm atm}$ from the same hadronic cascade and decay chain ($\pi^\pm, K^\pm \to \mu \to e$). Minor differences at low energies may arise from energy-dependent muon propagation effects before decay.

It is worth noting that the results presented here correspond to a single site (JUNO, at mid-latitude). The geomagnetic cutoff rigidity varies significantly with geographic location, and the fractional decomposition would differ for other detector sites. At equatorial locations with higher cutoff rigidities (e.g., TRIDENT in South China Sea), low-energy cosmic rays are more strongly suppressed, which would reduce the contribution from the lowest $E_{\rm cr}$ bins to the sub-GeV neutrino flux. At polar or near-polar sites (e.g., IceCube at the South Pole), where the geomagnetic cutoff is minimal, even the lowest-energy cosmic rays can reach the atmosphere, enhancing the low-$E_{\rm cr}$ contribution to low-energy neutrinos. At sufficiently high energies ($E_{\rm cr} \gtrsim 100$ GeV), however, the geomagnetic effects vanish and the decomposition becomes site-independent.

\subsubsection{Geomagnetic Modulation of the Azimuthal Distribution}

\begin{figure}[!t]
\begin{center}
\includegraphics[width=1.0\textwidth]{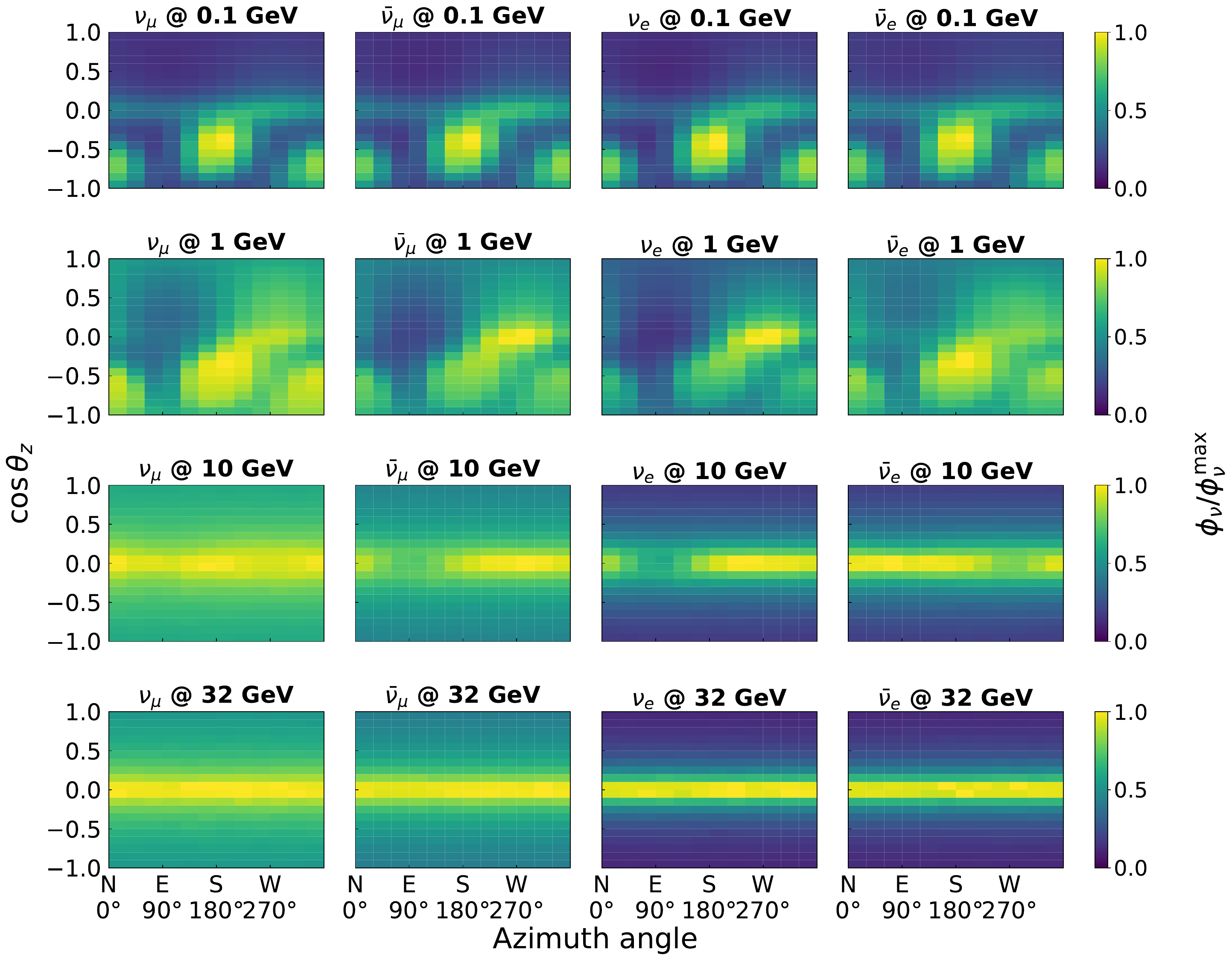}
\end{center}
\vspace{-0.8cm}
\caption{Normalized atmospheric neutrino flux at the JUNO site as a function of zenith angle ($\cos\theta_z$) and azimuth, for four flavors ($\nu_\mu$, $\bar{\nu}_\mu$, $\nu_e$, $\bar{\nu}_e$) at 0.1, 1, 10, and 32 GeV. Each panel is normalized to its own maximum flux, showing the relative distribution of zenith angle and azimuth within each flavor and energy bin.}
\label{fig:fluxtzvsphi}
\end{figure}

\begin{figure}[!t]
\begin{center}
\includegraphics[width=0.9\textwidth]{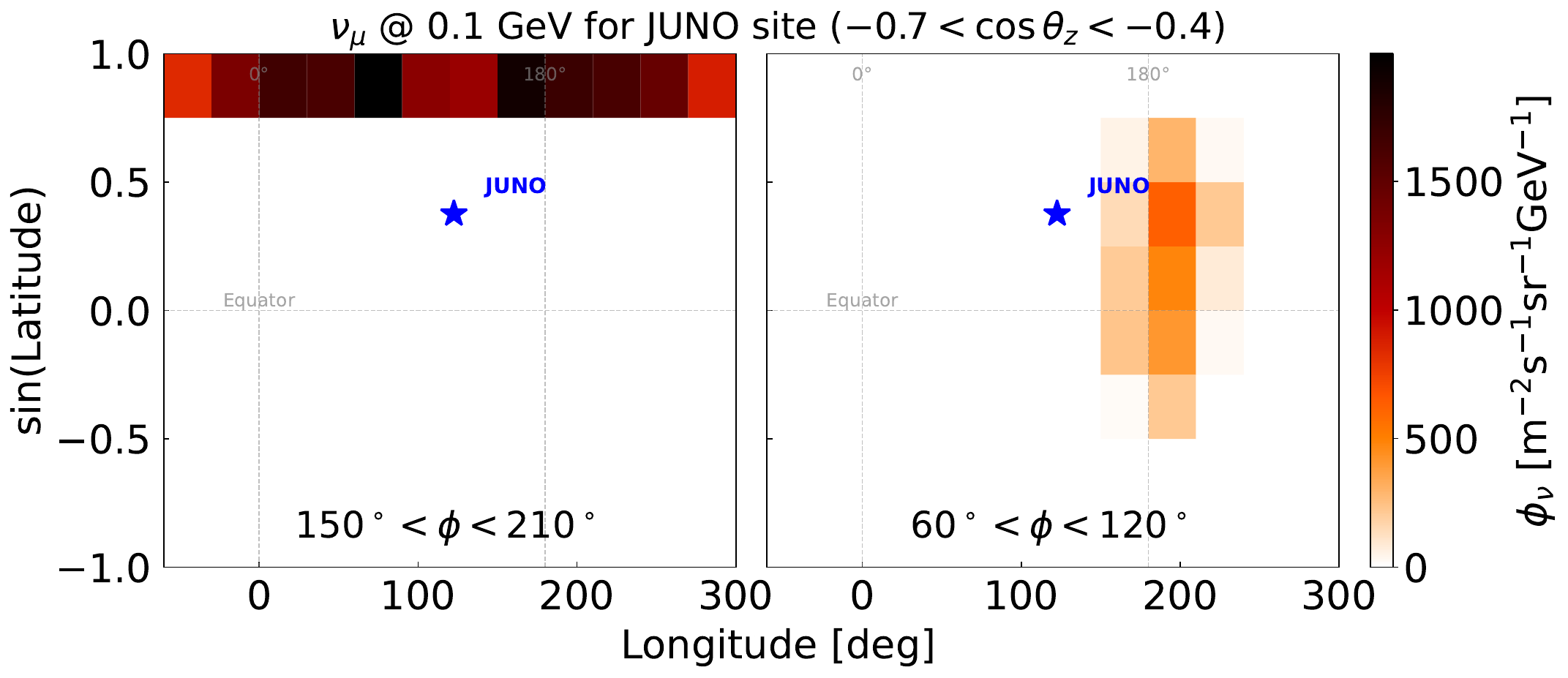}
\end{center}
\vspace{-0.8cm}
\caption{Flux-weighted production locations of 0.1 GeV atmospheric $\nu_\mu$ for the JUNO site in the zenith angle bin $-0.7 < \cos\theta_z < -0.4$, shown in geographic coordinates for two azimuth bins: $150^\circ < \phi< 210^\circ$ (left) and $60^\circ < \phi < 120^\circ$ (right). The star marker indicates the geographic position of the JUNO site.}
\label{fig:prod_loc2}
\end{figure}

\paragraph{Azimuthal modulation at low energies.}
We now examine how the geomagnetic field modulates the azimuthal distribution of these neutrinos.
Figure~\ref{fig:fluxtzvsphi} illustrates the angular distribution of the $\nu_{\rm atm}$ flux at the JUNO site across four energy bins (0.1, 1, 10, and 32 GeV) and four flavors. A consistent feature across all energies is the concentration of flux near the horizontal direction ($\cos\theta_z \approx 0$), which reflects the maximum atmospheric slant depth available for cosmic ray interactions along these trajectories. However, a distinct energy-dependent structure emerges in the azimuthal dimension. At 0.1 GeV within the zenith angle bin $-0.7 < \cos\theta_z < -0.4$, the flux exhibits significant azimuthal modulation, with variations of nearly a factor of two between different azimuth sectors (e.g., the excess around $150^\circ < \phi < 210^\circ$ compared to the deficit near $60^\circ < \phi < 120^\circ$). This modulation is present but noticeably weaker at 1 GeV and is virtually absent at 10 and 32 GeV, where the distribution becomes azimuthally uniform.

The origin of this azimuthal structure is elucidated in Fig.~\ref{fig:prod_loc2}, which maps the flux-weighted production locations of 0.1 GeV $\nu_\mu$ within the same zenith bin, separated by azimuth sector. The figure reveals a direct correlation between the arrival azimuth and the latitude of production. Neutrinos arriving from the high-flux azimuth sector ($150^\circ < \phi < 210^\circ$) are produced predominantly in high-latitude regions ($\sin(\text{Lat}) \approx 1.0$), whereas those from the low-flux sector ($60^\circ < \phi < 120^\circ$) originate near the equator and mid-latitudes. This geometric mapping demonstrates that the azimuthal modulation is a direct consequence of the Earth's magnetic field geometry: the high-latitude production zones correspond to regions with low geomagnetic cutoff rigidity, permitting greater access for low-energy cosmic rays, while equatorial production regions are shielded by high cutoff rigidities.

The disappearance of this modulation at higher neutrino energies is consistent with the primary cosmic ray energy decomposition presented in Fig.~\ref{fig:flux_ratio_by_Ecr}. At 0.1 GeV, the $\nu_{\rm atm}$ flux is dominated by contributions from low-energy primary cosmic rays ($E_{\rm cr} < 10$ GeV). These primaries have rigidities comparable to the geomagnetic cutoff, rendering the resulting neutrino flux highly sensitive to the latitude-dependent shielding shown in Fig.~\ref{fig:prod_loc2}. As the neutrino energy increases to 10 GeV and above, the flux becomes dominated by primaries with energies $E_{\rm cr} > 100$ GeV. These high-rigidity cosmic rays are essentially unaffected by the geomagnetic field regardless of latitude, resulting in the isotropic, azimuthally uniform flux distributions observed in the high-energy panels of Fig.~\ref{fig:fluxtzvsphi}.

\begin{figure}[!t]
\begin{center}
\includegraphics[width=0.7\textwidth]{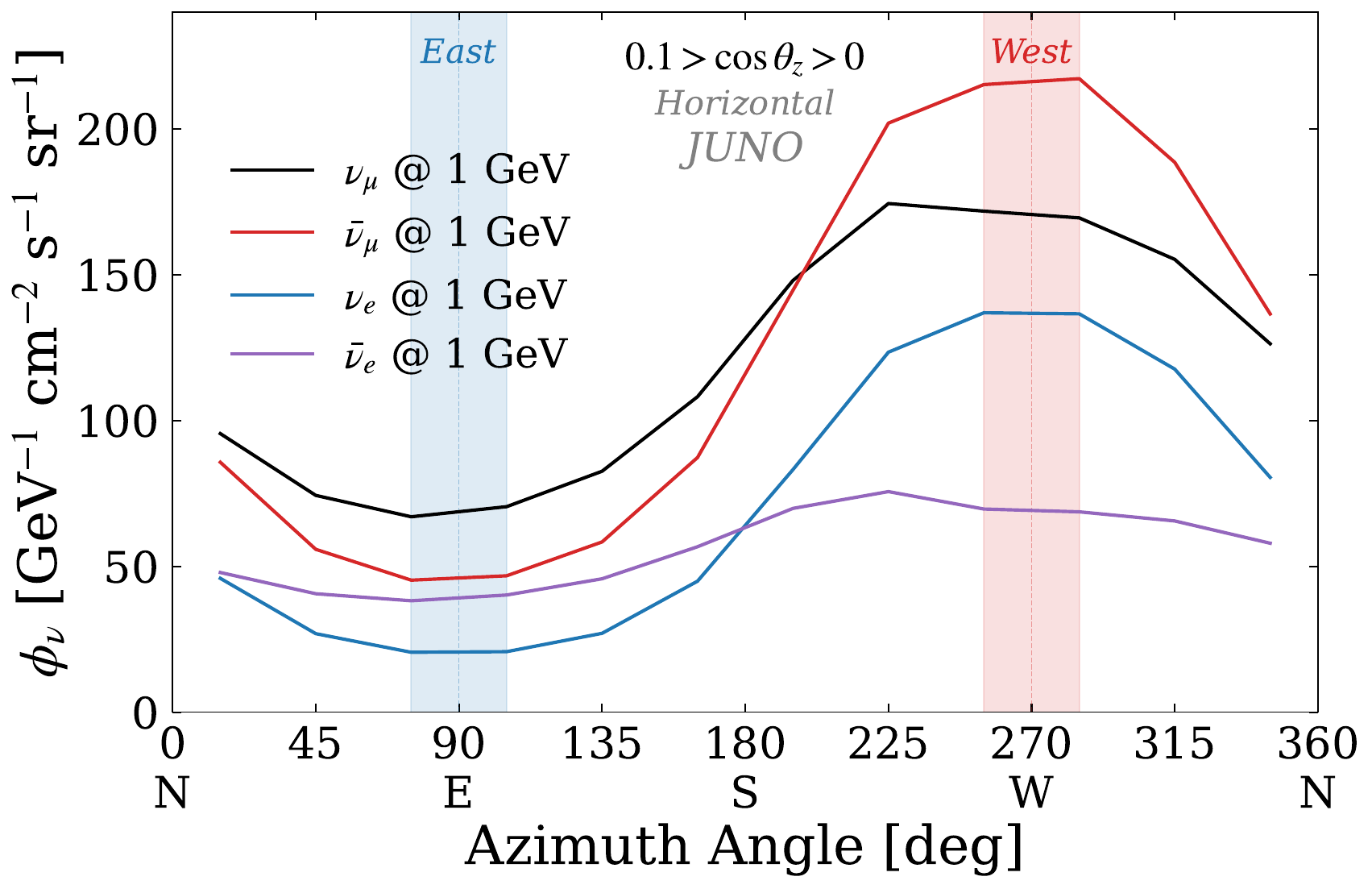}
\end{center}
\vspace{-0.8cm}
\caption{Azimuthal distribution of the atmospheric neutrino flux at 1 GeV for the horizontal zenith angle bin ($0 < \cos\theta_z < 0.1$) at the JUNO site. The shaded regions highlight the East and West sectors.}
\label{fig:EvsW}
\end{figure}

\paragraph{East-West effect.}
To further investigate the azimuthal structure at low energies, Figure~\ref{fig:EvsW} isolates the near-horizontal zenith angle bin at 1 GeV. The plot reveals a striking {\bf{East-West effect}}, where the neutrino flux peaks in the West sector ($270^\circ$) and drops to a minimum in the East ($90^\circ$). This pronounced asymmetry is a direct consequence of the Earth's magnetic field acting as a rigidity filter on charged cosmic rays. At the JUNO site (low geomagnetic latitude), primary cosmic rays arriving from the East must possess a higher rigidity to penetrate the magnetosphere compared to those arriving from the West. Since the primary cosmic ray spectrum is dominated by positively charged protons, this results in a suppressed flux of secondaries from the East and a significant enhancement from the West.

The asymmetry is further modulated by the geomagnetic bending of secondary muons in the atmosphere. Positive and negative muons ($\mu^+$ and $\mu^-$) are deflected in opposite directions before they decay, which introduces a flavor-dependent signature to the East-West effect. This is visible in Fig.~\ref{fig:EvsW}, where the asymmetry magnitude varies slightly between the muon-flavor and electron-flavor neutrinos, reflecting the distinct charge-sign dependence of their production and decay chains.

\subsubsection{Comparison with HKKMS15}\label{subsec:comparison_hkkms15}

\begin{figure}[!t]
\begin{center}
\includegraphics[width=0.9\textwidth]{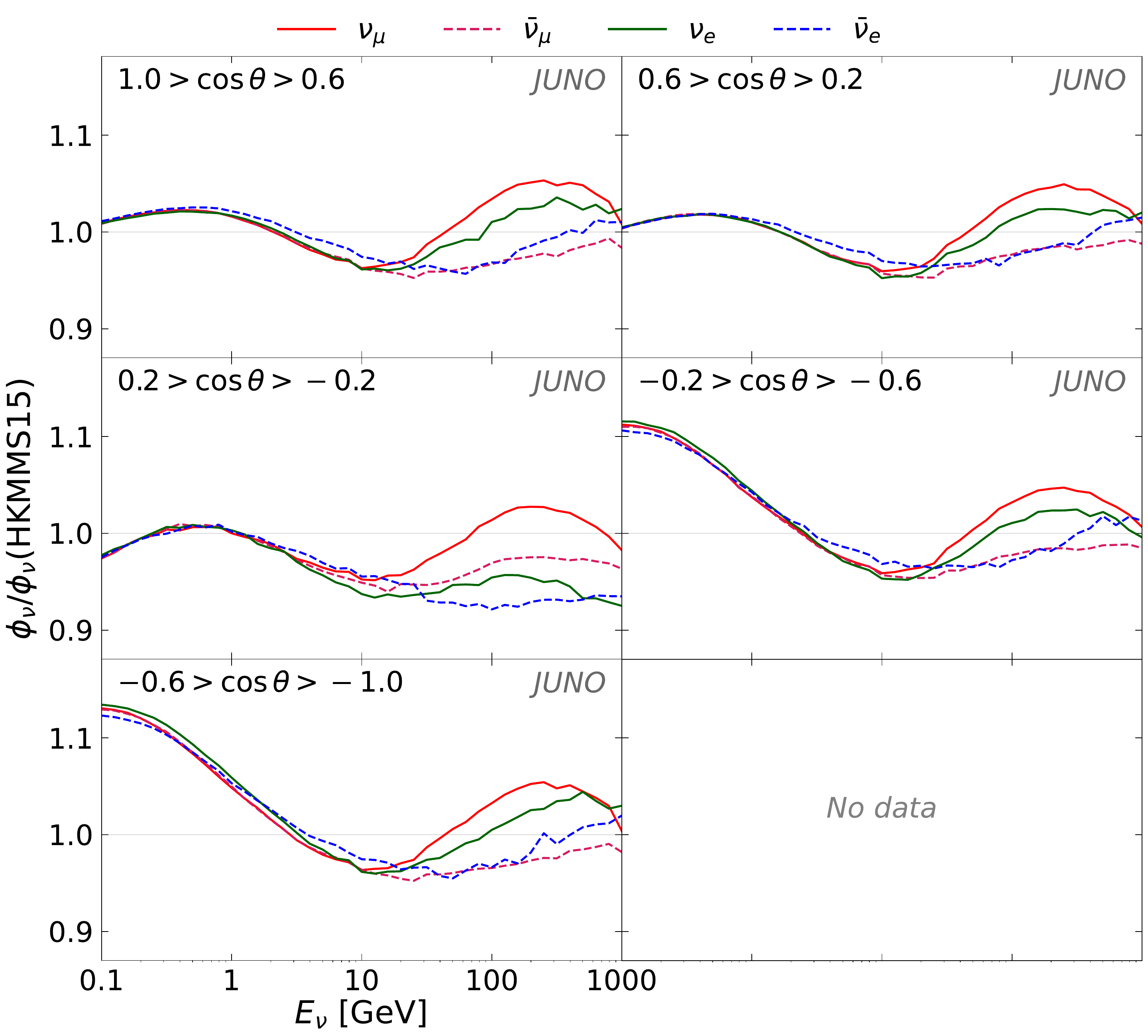}
\end{center}
\vspace{-0.8cm}
\caption{Ratio of the atmospheric neutrino flux calculated in the current work to that of the HKKMS15 model~\cite{Honda:2015fha}, $\phi_{\nu}/\phi_{\nu}(\rm HKKMS15)$, as a function of neutrino energy $E_\nu$ at the JUNO site. The panels are separated by zenith angle $\cos\theta_z$ bins: downward-going ($1.0 > \cos\theta_z > 0.6$, top-left), near-vertical downward ($0.6 > \cos\theta_z > 0.2$, top-right), near-horizontal ($0.2 > \cos\theta_z > -0.2$, middle-left), near-vertical upward ($-0.2 > \cos\theta_z > -0.6$, middle-right), and upward-going ($-0.6 > \cos\theta_z > -1.0$, bottom).}
\label{fig:vsH15}
\end{figure}

We now summarize the combined impact of the updated input models (AMS02-based primary cosmic ray model, IGRF2020 geomagnetic field, and muon-recalibrated hadronic interaction model) by comparing the current flux results with those of the HKKMS15 framework~\cite{Honda:2015fha} at the JUNO site. Figure~\ref{fig:vsH15} presents the ratio of the $\nu_{\rm atm}$ flux calculated in this work to that of HKKMS15, $\phi_{\nu}/\phi_{\nu}(\rm HKKMS15)$, as a function of neutrino energy $E_\nu$, for the four neutrino flavors ($\nu_\mu$, $\bar{\nu}_\mu$, $\nu_e$, $\bar{\nu}_e$)
and five zenith angle bins ranging from downward-going to upward-going directions. 

Below $\sim$10~GeV, the behavior of the ratio is qualitatively distinct from the high-energy regime. In this energy regime, all four flavor ratios deviate from unity in a nearly consistent manner, with relatively small splitting among flavors. This is because the deviations at these energies are driven primarily by the geomagnetic cutoff acting on the primary cosmic rays rather than by the charge-asymmetric details of the hadronic interaction model. The geomagnetic field filters incoming cosmic rays based on their rigidity, affecting all secondary particle species --- and hence all neutrino flavors --- in a similar way.

However, the magnitude of the deviation below $\sim$10~GeV exhibits a strong dependence on the zenith angle. The ratio rises most significantly for upward-going neutrinos, reaching deviations of several percent, while the enhancement for downward-going and horizontal neutrinos is comparatively modest. This striking zenith-angle dependence can be understood by considering the production locations of neutrinos arriving from different directions.

Upward-going neutrinos at the JUNO site predominantly originate from production regions on the opposite side of the Earth, which tend to have weaker horizontal geomagnetic field components and thus lower geomagnetic cutoff rigidities. In these low-cutoff regions, low-energy primary cosmic rays ($E_{\rm cr} \lesssim 10$~GeV) can more readily penetrate the atmosphere and produce secondary particles. Since the AMS02-based primary cosmic ray model has a higher flux than the AMS01-based model below $\sim$40~GeV, the enhanced low-energy cosmic ray flux in the AMS02-based model disproportionately benefits these low-cutoff production regions. As a result, the flux ratio for upward-going neutrinos shows a pronounced enhancement at low energies.

In contrast, downward-going neutrinos are produced locally near the detector site, where the geomagnetic cutoff rigidity is relatively high. The local high cutoff suppresses the access of low-energy cosmic rays regardless of which primary cosmic ray model is used, so the enhancement from the AMS02-based model is less pronounced. Horizontal neutrinos receive contributions from intermediate production locations, and their ratio exhibits a moderate deviation between the upward and downward extremes.

It is worth noting that the direct effect of updating the geomagnetic field model from IGRF2010 to IGRF2020 on the flux normalization is small, as the geomagnetic field acts globally and the change between consecutive IGRF versions is minor over a ten-year interval. The dominant effect of the geomagnetic field on the flux ratio in Fig.~\ref{fig:vsH15} arises indirectly through the coupling between the production-location-dependent cutoff rigidity and the spectral change of the primary cosmic ray model.

At high energies ($E_\nu \gtrsim 10$~GeV), the geomagnetic field effects become negligible for high-rigidity primary cosmic rays, and the ratios for all flavors approach values close to unity. However, a careful inspection reveals that the $\nu_\mu/\bar{\nu}_\mu$ pair and the $\nu_e/\bar{\nu}_e$ pair exhibit small but persistent splittings of order 1--3\% across all zenith angle bins. This flavor-dependent deviation at high energy is not caused by geomagnetic effects but
rather by the charge asymmetry inherent in the primary cosmic ray composition combined with the energy-dependent spectral changes between the AMS01-based and AMS02-based primary cosmic ray models and the recalibration of the hadronic interaction model.

Specifically, the primary cosmic ray flux is dominated by protons ($\sim$90\%), leading to an excess of $\pi^+$ over $\pi^-$ in hadronic interactions. This charge asymmetry propagates through the decay chain: $\pi^+ \rightarrow \mu^+ + \nu_\mu$ followed by
$\mu^+ \rightarrow e^+ + \nu_e + \bar{\nu}_\mu$, and
$\pi^- \rightarrow \mu^- + \bar{\nu}_\mu$ followed by
$\mu^- \rightarrow e^- + \bar{\nu}_e + \nu_\mu$.
The AMS02-based primary cosmic ray model, which is based on precise AMS02, BESS-polar, and PAMELA data, exhibits a different spectral shape compared to the AMS01-based model: higher flux below $\sim$40~GeV and lower flux above $\sim$40~GeV, with a maximum discrepancy of $\sim$50\% for primary protons. Because the $\pi^+/\pi^-$ and $K^+/K^-$ production ratios depend on the incident cosmic ray energy, this spectral change leads to a shift in the relative yields of positive and negative secondaries.

Furthermore, the hadronic interaction model has been recalibrated using the AMS02-based primary cosmic ray model and $\mu_{\rm atm}$ flux data. Since the $\mu_{\rm atm}$ charge ratio $\mu^+/\mu^-$ is approximately 1.2--1.3 (not unity), the recalibration affects $\mu^+$ and $\mu^-$ fluxes differently. Given that $\nu_\mu$ is primarily sourced from $\pi^+$ decay and $\mu^-$ decay, while $\bar{\nu}_\mu$ is sourced from $\pi^-$ decay and $\mu^+$ decay, the differential corrections to $\mu^+$ and $\mu^-$ translate into different normalization shifts for $\nu_\mu$ and $\bar{\nu}_\mu$. The same mechanism, acting through the second-generation decays, produces the $\nu_e/\bar{\nu}_e$ splitting. At energies above $\sim$10~GeV, the contribution from kaon decays also becomes significant, and the $K^+/K^-$ production asymmetry further amplifies the flavor-dependent differences. These effects are all physically expected and are consistent with the intrinsic charge asymmetry of the $\nu_{\rm atm}$ production process.

In summary, the comparison with HKKMS15 confirms that the current calculation is consistent with the established framework at high energies while providing improved precision at low energies through updated input models. The deviations observed across the energy range of Fig.~\ref{fig:vsH15} are within the estimated systematic uncertainties of the HKKMS flux model ($\sim$10\% for $E_\nu > 1$~GeV and $\sim$15--25\% for $E_\nu < 1$~GeV) and are physically well-understood. It should be noted that the energy range below $\sim$100~MeV, where the inclusion of neutrinos from $\mu_{\rm atm}$ propagation inside the Earth leads to a qualitatively new contribution absent in HKKMS15, is not covered in Fig.~\ref{fig:vsH15} and is discussed separately in Sec.~\ref{subsec:elow100}.

\subsection{Region of $E_\nu < 100$ MeV}\label{subsec:elow100}

\begin{figure}[!t]
\begin{center}
\includegraphics[width=0.7\textwidth]{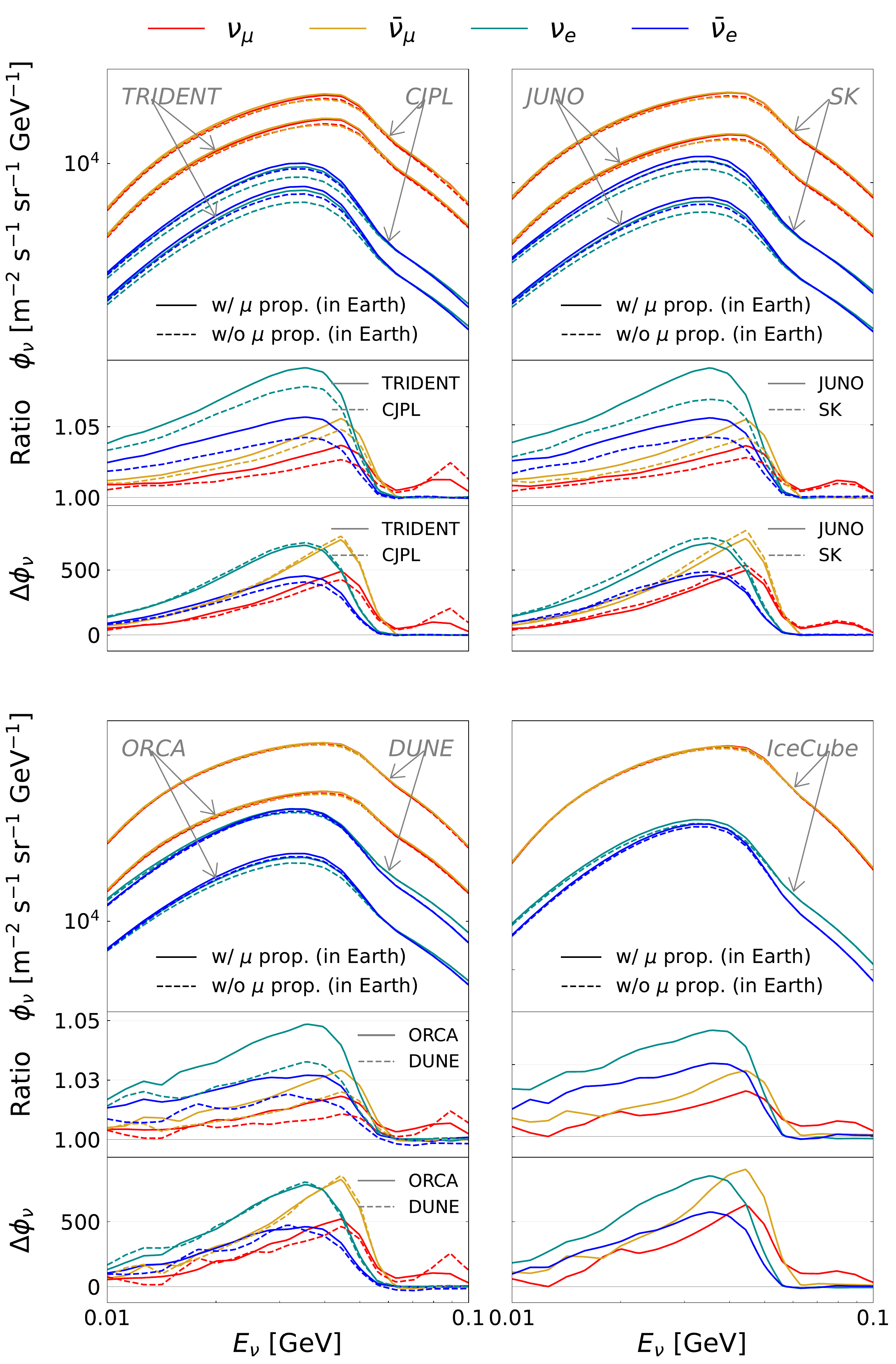}
\end{center}
\vspace{-0.8cm}
\caption{All-direction-averaged atmospheric neutrino fluxes for seven experiment sites in the energy range $E_\nu < 0.1$ GeV, arranged in four panels: (top-left) TRIDENT and CJPL, (top-right) JUNO and SK, (bottom-left) KM3NeT/ORCA and DUNE, and (bottom-right) IceCube. Within each panel, the top row shows the absolute flux $\phi_\nu$ with (solid lines) and without (dashed lines) muon propagation inside the Earth; site labels with arrows distinguish the two overlaid data sets (each containing four neutrino flavors, with and without muon propagation). The middle row displays the ratio $\phi_\nu^{\rm (w/)}/\phi_\nu^{\rm (w/o)}$ of the two calculations, highlighting the relative enhancement due to muon propagation. The bottom row shows their absolute difference $\Delta\phi_\nu \equiv \phi_\nu^{\rm (w/)} - \phi_\nu^{\rm (w/o)}$.}
\label{fig:mu_prop_comparison}
\end{figure}

\begin{figure}[!t]
\begin{center}
\includegraphics[width=1.0\textwidth]{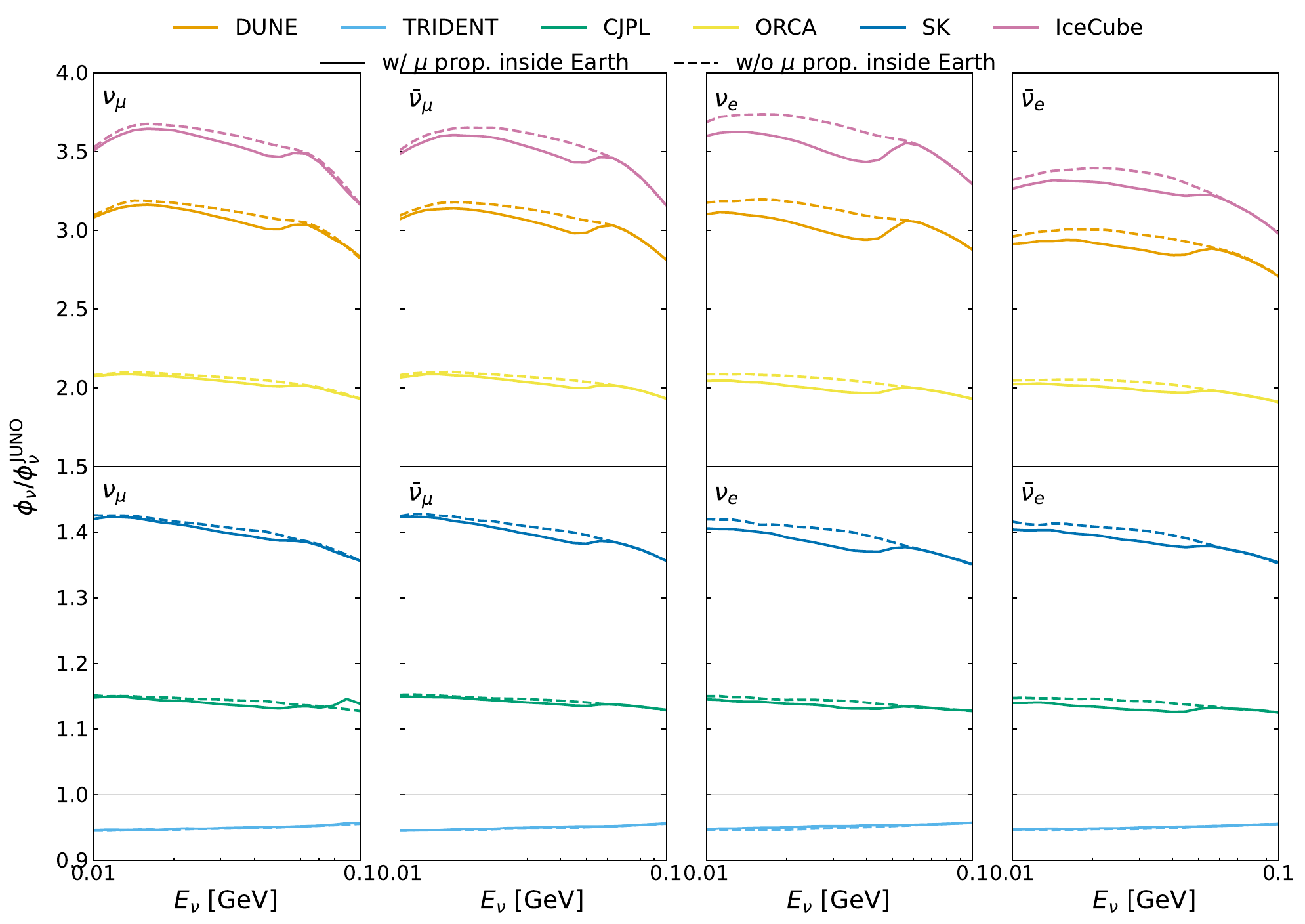}
\end{center}
\vspace{-0.8cm}
\caption{Comparison of the all-direction-averaged atmospheric neutrino fluxes across six experiment sites (DUNE, TRIDENT, CJPL, KM3NeT/ORCA, SK, and IceCube) relative to the JUNO reference site, with JUNO set to unity.}
\label{fig:site_comparison}
\end{figure}

In Sec.~\ref{subsec:comparison_hkkms15}, we compare the current flux results with the HKKMS15 calculation for $E_\nu > 100$~MeV, where the updated primary cosmic ray model (AMS02-based vs.~AMS01-based) and the recalibrated hadronic interaction model produce energy-dependent deviations of 2--10\% across all flavors and sites. However, the energy range below $\sim$100~MeV---where the inclusion of neutrinos from $\mu_{\rm atm}$ propagation inside the Earth leads to a qualitatively new contribution---is not covered in Fig.~\ref{fig:vsH15} and is discussed separately in this section.

The scientific motivation for precise $\nu_{\rm atm}$ flux calculations in this energy range is twofold. First, the DSNB signal is expected to peak at 10--30~MeV, where $\nu_{\rm atm}$ constitute a dominant irreducible background. Second, indirect dark matter searches targeting low-mass WIMPs also operate in the tens of MeV regime. Previous 3D flux calculations (HKKMS11~\cite{Honda:2011nf}, HKKMS15~\cite{Honda:2015fha}, Bartol~\cite{Barr:2004br}, FLUKA~\cite{Battistoni:2002ew}) did not account for the contribution from $\mu_{\rm atm}$ that propagate inside the Earth, stop, and subsequently decay or are captured by nuclei.

In the following, we first establish the \textit{baseline flux} calculated \textit{without} muon propagation inside the Earth (equivalent to the HKKMS15 approach), then add the \textit{muon-propagation contribution} from muons that stop inside the Earth and subsequently decay or undergo nuclear capture.

The top row of Fig.~\ref{fig:mu_prop_comparison} (dashed lines) shows the all-direction-averaged $\nu_{\rm atm}$ fluxes for seven experiment sites (TRIDENT, JUNO, CJPL, SK, KM3NeT/ORCA, DUNE, and IceCube) calculated \textit{without} muon propagation inside the Earth---i.e., the baseline flux. The spectral shapes are similar across sites for each flavor, but the absolute flux levels differ dramatically. Figure~\ref{fig:site_comparison} (dashed curves) quantifies these site-to-site differences relative to the JUNO reference site: IceCube receives $\sim$3.7 times the JUNO flux for $\nu_\mu$ and $\bar{\nu}_\mu$, DUNE $\sim$3.1 times, KM3NeT/ORCA $\sim$2.1 times, SK $\sim$1.4 times, and CJPL $\sim$1.15 times, while TRIDENT (near the geomagnetic equator) is slightly below JUNO at $\sim$0.98 times. These differences are entirely driven by geomagnetic rigidity cutoff effects---the horizontal geomagnetic field component $B_h$ ranges from $\sim$180~nT at IceCube to $\sim$39\,449~nT at TRIDENT (see Fig.~\ref{fig:geomagnetic_field}). This constitutes the baseline against which the muon-propagation contribution is evaluated.

When muon propagation inside the Earth is included (solid lines in the top row of Fig.~\ref{fig:mu_prop_comparison}), the flux begins to deviate from the baseline below $\sim$100~MeV, and the effect grows as the energy decreases. The bottom row of Fig.~\ref{fig:mu_prop_comparison} displays the absolute difference $\phi_\nu^{\rm (w/)} - \phi_\nu^{\rm (w/o)}$ for each site, revealing three key features.

\textbf{First, the muon-propagation contribution is a global effect.} The difference curves for all seven sites are of comparable magnitude and share a similar shape, with a peak in the $\sim$10--50~MeV range. Because the Earth is effectively transparent to neutrinos, any neutrino produced by muon decay or capture within the Earth can reach the detector regardless of where the muon stopped. The microscopic processes governing muon energy loss, stopping, decay, and nuclear capture in the Earth's crust or ice layer operate as an approximately site-independent mechanism, producing a globally consistent absolute contribution.

\textbf{Second, the difference curves are not perfectly identical across sites.} The residual variations arise from site-specific geometric factors, including the sky coverage of observable arrival directions, the angular distribution of muons entering the Earth, and the local medium composition (rock vs.~water, as determined by the CRUST1.0 model). The medium composition affects the muon energy loss rate and the nuclear capture probability. Local mountain profile effects are not yet included in the current calculation and will be addressed in future work.

\textbf{Third, the flavor hierarchy follows $\nu_e \approx \bar{\nu}_\mu > \nu_\mu > \bar{\nu}_e$.} This ordering is primarily driven by the charge asymmetry of the $\mu_{\rm atm}$ that stop inside the Earth. The muons contributing to this process originate from a broad energy range of primary cosmic rays, where the proton-dominated composition leads to an excess of $\pi^+$ over $\pi^-$ and, consequently, $\mu^+$ over $\mu^-$. Although the $\mu^+/\mu^-$ ratio approaches unity at very low energies, the stopping muon sample includes higher-energy muons for which $\mu^+ > \mu^-$. As a result, the free decay of $\mu^+$ ($\mu^+ \to e^+ + \nu_e + \bar{\nu}_\mu$) produces more neutrinos than the decay of $\mu^-$. This makes $\nu_e$ and $\bar{\nu}_\mu$ the dominant flavors in the muon-propagation contribution. The hierarchy between $\nu_\mu$ and $\bar{\nu}_e$ is further shaped by nuclear capture: the capture of $\mu^-$ suppresses $\bar{\nu}_e$ production while generating additional $\nu_\mu$, leading to $\nu_\mu > \bar{\nu}_e$.

The $\nu_\mu$ contribution exhibits a distinctive high-energy feature. The neutrinos from nuclear capture are not monoenergetic but rather have a structured energy distribution comprising a low-energy continuum component and a high-energy peak at $\sim$100~MeV. In water-dominated environments (oxygen), the low-energy component dominates; in rock-dominated environments (silicon), the high-energy peak is more pronounced. Consequently, rock-based sites exhibit a significant nuclear-capture contribution above 60~MeV, resulting in a noticeable high-energy tail in the $\nu_\mu$ difference curve. For nuclei heavier than silicon, the silicon energy spectrum is currently used as an approximation and will be updated in future work. The $\nu_\mu$ from in-atom $\mu^-$ decay follows a continuous Michel spectrum with an endpoint at $m_\mu/2 \approx 52.8$~MeV and a peak at $\sim$30--40~MeV.

The middle row of Fig.~\ref{fig:mu_prop_comparison} shows the ratio $\phi_\nu^{\rm (w/)} / \phi_\nu^{\rm (w/o)}$, which reveals a markedly different pattern from the absolute difference. The relative enhancement varies significantly across sites: IceCube shows the smallest ratio ($\sim$1.02 for $\nu_\mu$), while TRIDENT shows the largest ($\sim$1.07--1.08). This is a direct consequence of the global nature of the muon-propagation contribution combined with the site-dependent baseline flux. Since the absolute increase is approximately constant across sites, the \textit{fractional} increase is inversely proportional to the baseline flux: sites with weaker geomagnetic cutoff (larger baseline) exhibit smaller relative enhancements, and vice versa. In other words, the geomagnetic field ``dilutes'' the relative contribution of muon propagation at high-flux sites while amplifying it at low-flux sites.
This behavior can be summarized as:
\begin{equation}
\frac{\phi^{\rm (w/)}}{\phi^{\rm (w/o)}} \approx 1 + \frac{\Delta_{\rm global}}{\phi^{\rm (w/o)}_{\rm site}},
\end{equation}
where $\Delta_{\rm global}$ is the approximately site-independent absolute contribution from muon propagation, and $\phi^{\rm (w/o)}_{\rm site}$ is the baseline flux modulated by the local geomagnetic rigidity cutoff.

Figure~\ref{fig:site_comparison} provides an independent cross-check of the above interpretation. Each panel compares the flux ratio $\phi_\nu^{\rm (site)} / \phi_\nu^{\rm (JUNO)}$ with and without muon propagation. The top row groups sites with fluxes significantly above JUNO (IceCube, DUNE, KM3NeT/ORCA), while the bottom row groups sites with fluxes closer to JUNO (SK, CJPL, TRIDENT).

In the bottom-row panels, the solid and dashed curves nearly overlap, reflecting the fact that the baseline fluxes of these sites are comparable to JUNO, so the relative increases from muon propagation are similar and the ratios to JUNO change minimally. In contrast, the top-row panels show that the solid curves lie slightly below the dashed ones. This is the expected behavior of an \textit{additive} global correction: JUNO has a lower baseline flux than IceCube, DUNE, or KM3NeT/ORCA, so the \textit{fractional} increase from muon propagation is larger at JUNO than at these high-flux sites, causing their ratios to JUNO to decrease slightly when muon propagation is included.

This observation confirms two points. First, the absolute contribution from muon propagation is globally consistent across sites, as evidenced by the similar difference curves in Fig.~\ref{fig:mu_prop_comparison}. Second, the impact of this global correction on inter-site flux ratios depends on the choice of normalization baseline---when the baseline site has a lower flux, the ratios to it decrease; when the baseline is comparable, the ratios remain essentially unchanged. The substantial site-to-site flux variations (up to a factor of $\sim$4 between IceCube and TRIDENT) are therefore dominated by geomagnetic effects, not by the muon-propagation mechanism.

\medskip
\noindent
\textbf{Implications for DSNB and dark matter searches.} The precise flux predictions presented in this section provide essential inputs for ongoing and upcoming rare-event searches. For the DSNB search at JUNO, the $\nu_{\rm atm}$ background in the 10--30~MeV window is now quantified with the inclusion of muon-propagation contributions, which were previously neglected in 3D flux calculations. Similarly, for indirect dark matter searches targeting low-mass WIMPs in the tens-of-MeV range, the updated flux predictions reduce the systematic uncertainty in the $\nu_{\rm atm}$ background estimate. The site-by-site comparison also enables cross-experiment calibration: experiments at different geomagnetic latitudes can now use consistent flux models to interpret their low-energy neutrino observations.

\section{Flux Model Uncertainty Estimation}
\label{sec:uncertainty}

The systematic uncertainty of the $\nu_{\rm atm}$ flux is dominated by two sources: primary cosmic ray spectra and hadronic interaction models. Recent precision measurements by AMS02 and other experiments~\cite{AMS:2015tnn,AMS:2015azc, Abe:2015mga, PAMELA:2011mvy} have reduced the primary cosmic ray uncertainty to a few percent. In this section, we focus on updating the hadronic uncertainty estimate using the atmospheric muon constraint method, which provides the most significant improvement over previous calculations.

\textbf{Site applicability.} The atmospheric muon constraint method applies to all sites. Site-specific differences may arise only at low energies ($E_\nu < 1$ GeV) where geomagnetic effects modulate the primary cosmic ray composition differently for different sites.

\textbf{Methodology.} In previous HKKMS analyses~\cite{Honda:2006qj, Sanuki:2006yd}, the total systematic uncertainty was estimated as a quadrature sum:
\begin{equation}
\delta^2_{\rm tot}(E_\nu) = \delta^2_{\rm had}(E_\nu) + \delta^2_{\sigma}(E_\nu) + \delta^2_{\rm air}(E_\nu),
\label{eq:uncertainty}
\end{equation}
where $\delta_{\rm had}$ represents the hadronic production uncertainty, $\delta_{\sigma}$ the cross-section uncertainty, and $\delta_{\rm air}$ the atmospheric model uncertainty. For $E_\nu \gtrsim 1$ GeV, $\delta_{\rm had}$ was obtained by comparing flux calculations with $\mu_{\rm atm}$ measurements. However, this approach becomes invalid below 1 GeV, where muon energy loss in the atmosphere deforms the phase-space correlation between muons and neutrinos.

In this work, we update $\delta_{\rm had}$ using the quantitative atmospheric muon constraint method developed by Honda \textit{et al.}~\cite{Honda:2019ymh}. This method generates $\sim 3 \times 10^6$ model variations of the hadronic interaction and constrains them with precisely measured $\mu_{\rm atm}$ flux data, yielding the irreducible uncertainty $\zeta_0$---the component that cannot be eliminated even with perfect muon reconstruction.

\begin{figure}[!t]
\begin{center}
\includegraphics[width=1.0\textwidth]{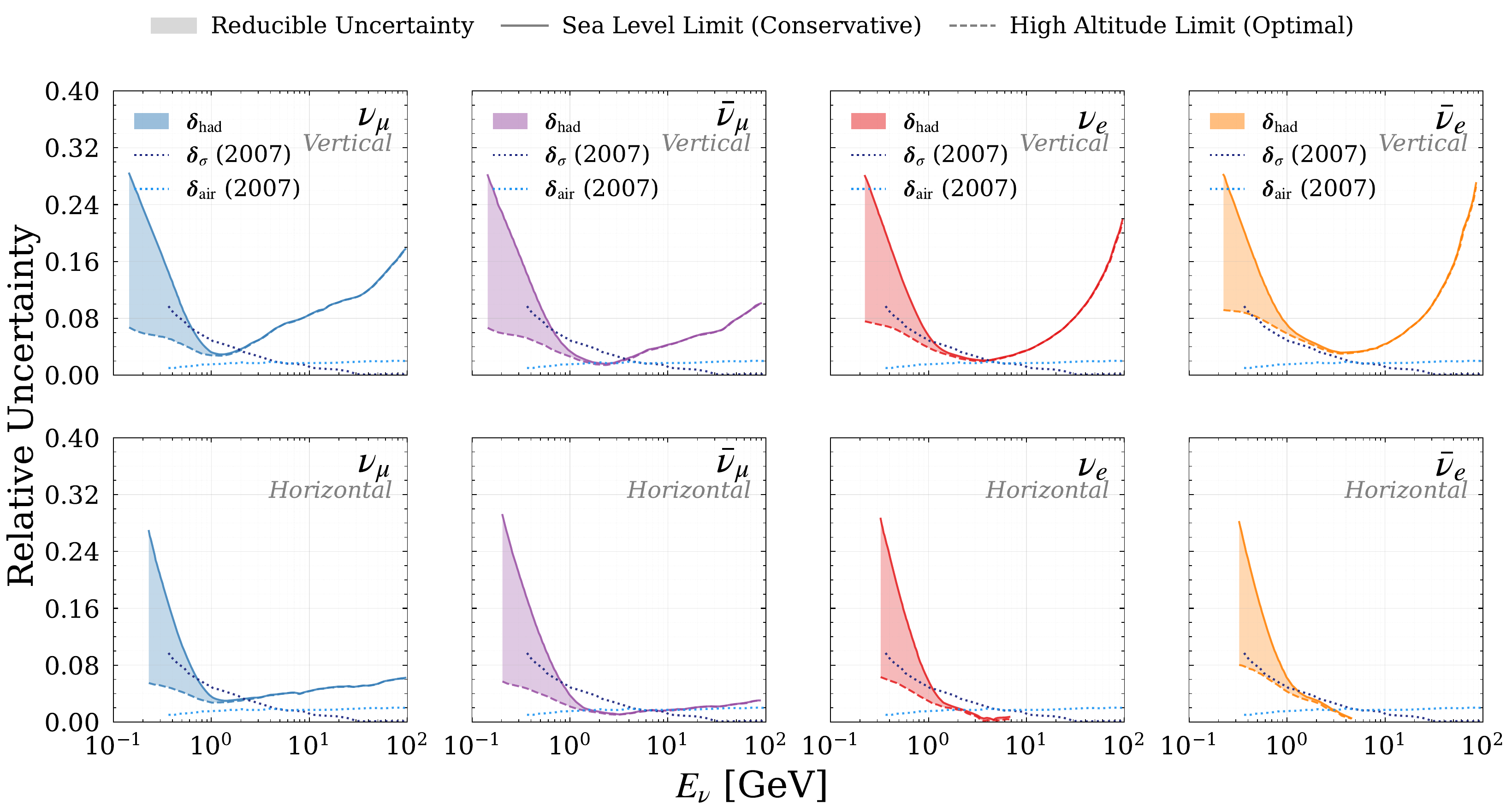}
\end{center}
\vspace{-0.8cm}
\caption{Breakdown of the hadronic uncertainty component ($\delta_{\rm had}$) in the atmospheric neutrino flux calculation, derived from the atmospheric muon constraint method~\cite{Honda:2019ymh}. The figure is organized by direction (Vertical $\cos\theta_z\sim1$ in the top row, Horizontal $\cos\theta_z\sim0$ in the bottom row) and flavor ($\nu_\mu, \bar{\nu}_\mu, \nu_e, \bar{\nu}_e$). The \textbf{shaded bands} represent the reducible uncertainty range. The \textbf{upper 
solid boundary} corresponds to the conservative constraint from sea-level muon observations (corresponding to Fig.~9 of Ref.~\cite{Honda:2019ymh}), while the \textbf{lower dashed 
boundary} corresponds to the optimal constraint from high-altitude observations at Hanle, 4500~m a.s.l. (Fig.~12 of Ref.~\cite{Honda:2019ymh}). The \textbf{dotted lines} represent the cross-section ($\delta_\sigma$) and atmospheric density model ($\delta_{\rm air}$) uncertainties inherited from the HKKMS06 analysis~\cite{Honda:2006qj}.}
\label{fig:figunc1}
\end{figure}

\textbf{Key results.} Figure~\ref{fig:figunc1} presents the hadronic uncertainty $\delta_{\rm had}$ as a function of neutrino energy. The following features emerge:

\begin{itemize}
\item \textbf{Significant improvement below 1~GeV.} Across 0.1--1~GeV, the sea-level constraint yields $\delta_{\rm had} \approx 3$--29\% for $\nu_\mu$/$\bar{\nu}_\mu$ (from $\sim 29\%$ at the lowest accessible energy of $\sim 0.15$~GeV to $\sim 3\%$ at 1~GeV); for $\nu_e$/$\bar{\nu}_e$ the sea-level data start at $\sim 0.2$~GeV with $\delta_{\rm had} \approx 28\%$, declining to $\sim 5\%$ at 1~GeV. The high-altitude constraint, which extends down to $\sim 0.1$~GeV for all flavors, reduces the $\nu_\mu$/$\bar{\nu}_\mu$ range to $\sim 3$--8\% and $\nu_e$/$\bar{\nu}_e$ to $\sim 4$--10\%.

\item \textbf{Directional dependence.} The horizontal direction consistently yields lower uncertainties than the vertical direction. For $\nu_\mu$ in the horizontal direction, the optimal constraint gives $\delta_{\rm had} \approx 3\%$ at 1~GeV, rising only gradually to $\sim 4$--5\% at 10~GeV. For $\nu_\mu$ in the vertical direction, the corresponding values are $\sim 3\%$ at 1~GeV and $\sim 8\%$ at 10~GeV. This difference reflects the longer atmospheric path of horizontal trajectories, which better preserves the phase-space correlation between hadronic interactions and observed muons.

\item \textbf{Flavor dependence at high energy.} Above $\sim 10$~GeV in the vertical direction, $\delta_{\rm had}$ rises more steeply for $\nu_e$ and $\bar{\nu}_e$ than for $\nu_\mu$ and $\bar{\nu}_\mu$. This arises because electron neutrinos are produced exclusively through secondary muon decay: as the muon decay probability decreases with energy, the phase-space link between surface muon observations and $\nu_e$ production is progressively weakened.
\end{itemize}

The rise of $\delta_{\rm had}$ for all flavors above a few GeV also reflects increasing kaon production contributions, which are less constrained by surface muon data than pion production. The $\delta_{\sigma}$ (cross-section) and $\delta_{\rm air}$ (atmospheric model) uncertainties, inherited from HKKMS06~\cite{Honda:2006qj}, vary with energy: $\delta_{\sigma}$ decreases from $\sim 10\%$ at $\sim 0.35$~GeV to below 2\% above $\sim 10$~GeV, while $\delta_{\rm air}$ remains at the $\sim 1$--3\% level across the energy range.

It is instructive to compare the muon-constrained uncertainties with those obtained from an independent methodology. Sato et al.~\cite{Sato:2026avb} recently employed accelerator data (notably from NA61/SHINE) to directly tune hadronic interaction models for low-energy atmospheric neutrino fluxes, reporting uncertainties of 7--9\% for $E_\nu < 1$~GeV and 5--7\% for $1 < E_\nu < 10$~GeV. The broad consistency between these two independent approaches---muon-based and accelerator-based---provides confidence in the estimated uncertainty range.

\begin{figure}[!t]
\begin{center}
\includegraphics[width=1.0\textwidth]{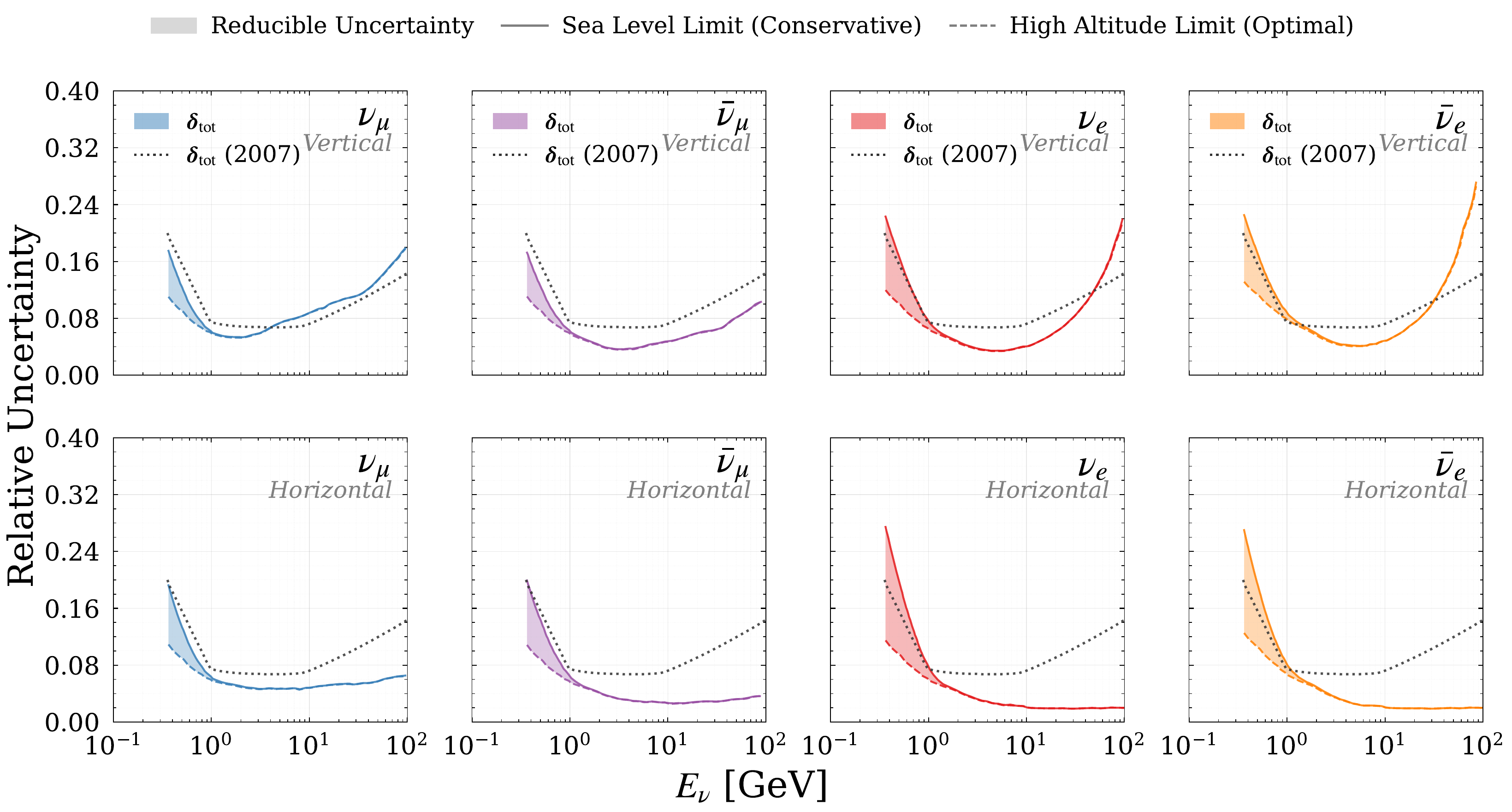}
\end{center}
\vspace{-0.8cm}
\caption{Total relative uncertainty ($\delta_{\rm tot}$) of the atmospheric neutrino flux. The layout follows Fig.~\ref{fig:figunc1}.
The \textbf{shaded bands} show the updated total uncertainty calculated in this work, combining the muon-constrained hadronic uncertainty with $\delta_\sigma$ and $\delta_{\rm air}$. The band width reflects the range between the conservative (sea-level) 
and optimal (high-altitude) scenarios.
The \textbf{dotted black lines} show the historical total uncertainty estimates ($\delta_{\rm tot}(2007)$) from the HKKMS06 framework ~\cite{Honda:2006qj} for reference.}
\label{fig:totunc}
\end{figure}

Figure~\ref{fig:totunc} shows the total uncertainty $\delta_{\rm tot}$, combining the updated $\delta_{\rm had}$ with $\delta_{\sigma}$ and $\delta_{\rm air}$. The displayed energy range is $\sim 0.35$--100~GeV, limited by the $\delta_{\sigma}$ and $\delta_{\rm air}$ input data which start at $\sim 0.35$~GeV. In the region where $\delta_{\rm had}$ is minimized (horizontal direction, 1--10~GeV), $\delta_{\rm tot}$ is reduced to $\sim 3$--6\% for all flavors, compared to $\sim 7\%$ at the same energies in the HKKMS06 analysis~\cite{Honda:2006qj} (whose total uncertainty spans $\sim 7$--25\% over the $\sim 0.35$~GeV--1~TeV range). From 0.35 to 1~GeV, the updated $\delta_{\rm tot}$ ranges from $\sim 6$--23\% (sea-level constraint) to $\sim 6$--12\% (high-altitude constraint), representing a significant improvement over the previous $\sim 7$--20\%. Above 10~GeV, $\delta_{\rm tot}$ rises with energy, driven by the increasing $\delta_{\rm had}$ from kaon contributions and muon decay suppression; for $\nu_e$/$\bar{\nu}_e$ in the vertical direction, the rise is steeper, reaching $\sim 15$--22\% at 100~GeV under the both constraints.

\section{Future Considerations and Refinements}\label{sec:future}

In the preceding sections, we have described the global models that form the foundation of the current 3D $\nu_{\rm atm}$ flux calculation and presented the resulting flux predictions and uncertainty estimates. While these results provide a robust baseline for the full energy spectrum, further improvements in precision can be achieved by incorporating site-specific environmental details and updating to the latest experimental constraints. In this section, we outline several refinements---ranging from local terrain and magnetic field corrections to time-dependent solar modulation and updated atmospheric and hadronic interaction models---that would further improve the flux predictions, particularly in low-to-intermediate energy regions. Since these refinements target specific physical regimes or reduce global systematic uncertainties, the core results presented in this paper---which span from 10~MeV to $10^4$~GeV---remain robust.
\begin{description}
    \item[Local mountain profile (relevant for $E_\nu \lesssim 100$~MeV).] The mountain topography surrounding the detector site modifies the local overburden through which $\mu_{\rm atm}$ propagate after entering the Earth. As discussed in Appendix~\ref{app:muon}, muons that penetrate the Earth lose energy via electromagnetic interactions, eventually stop, and may be captured by atomic nuclei or decay in orbit, producing low-energy neutrinos. The detailed path length and material composition along each muon trajectory depend on the local terrain, which can enhance or reduce the stopping probability for muons arriving from specific directions.
    Since the neutrinos produced via this mechanism are predominantly below 100~MeV, the mountain profile effect is confined to this energy region. High-resolution terrain data provided by the JUNO and CJPL collaborations are available for incorporation into the simulation. For other sites, global digital elevation models such as SRTM~\cite{SRTM1} or ASTER GDEM~\cite{ASTER1} (30~m resolution) provide comprehensive coverage and can serve as reliable alternatives. We plan to incorporate these terrain profiles into the muon propagation module in a forthcoming study.

    \item[Solar modulation (relevant for $E_\nu \lesssim 1$~GeV, especially below 100~MeV).] The 11-year solar cycle introduces time-varying suppression of low-energy cosmic rays entering the heliosphere, directly impacting the $\nu_{\rm atm}$ flux below $\sim 1$~GeV, and is most pronounced in the sub-100~MeV region where dark matter and DSNB searches operate. Rather than relying solely on the conventional force-field approximation, which utilizes simplified theoretical models, we plan to construct a time-dependent solar modulation model directly based on the precise, annual cosmic ray measurements from AMS02~\cite{AMS:2021qln, AMS:2022ojy, AMS:2025pgu} and BESS-polar~\cite{Abe:2015mga}. These time-series data provide rigorous empirical constraints on the modulated spectra over multiple solar cycles. By directly incorporating these measurements, we can establish a more reliable and realistic primary cosmic ray flux model for the low-energy range, thereby reducing the systematic uncertainties associated with older, purely theoretical parameterizations.
    \item[Local geomagnetic field (relevant for $E_\nu \lesssim 10$~GeV).] The global IGRF2020 model captures the large-scale geomagnetic structure but does not resolve crustal magnetic anomalies that may introduce sub-percent corrections to the local rigidity cutoff. Such corrections primarily affect cosmic rays at lower rigidities ($R_{\rm cr} \lesssim 10$~GV), corresponding to $\nu_{\rm atm}$ below $\sim 10$~GeV. Global magnetic anomaly grids such as EMAG2~\cite{EMAG2} and the WDMAM~\cite{WDMAM} are publicly available and could, in principle, be superimposed on the IGRF field to provide a higher-resolution description of the geomagnetic environment at each detector site. However, these models are derived primarily from satellite and airborne measurements at altitudes of several kilometers, and their precision at ground level---particularly in oceanic regions (e.g., TRIDENT, KM3NeT/ORCA)---remains limited. Site-specific ground-based magnetic surveys would be required to fully resolve local anomalies, but such data have not been published for any of the detector sites considered in this work. In future studies, we will evaluate the feasibility of incorporating EMAG2 or WDMAM into our backward trajectory tracing framework as an interim refinement.

    \item[Atmospheric model (relevant for the full energy spectrum).] The NRLMSISE-00 model used in this work provides a site- and time-dependent atmospheric description, which already accounts for seasonal variations. However, as a global empirical model, it represents a smoothed average of atmospheric conditions. Local meteorological data (e.g., from radiosondes) may reveal deviations from this model---such as specific weather patterns or short-term density fluctuations---which can modify the neutrino production height distribution and the competition between meson decay and re-interaction. These effects are most relevant for neutrinos produced by hadronic cascades at intermediate altitudes, which dominate the flux up to $\sim 100$~GeV. Above this energy, the production process becomes increasingly insensitive to atmospheric details. Access to local radiosonde or meteorological station data would enable a quantitative assessment of these model-vs-reality deviations. Furthermore, we plan to update the atmospheric density profile to the newer NRLMSISE 2.0 model~\cite{NRLMSISE20} in future work, which incorporates improved data and physics descriptions over the NRLMSISE-00 version. Until such updates are obtained, the current calculation relies on the NRLMSISE-00 predictions, which are expected to capture the dominant contribution.
    \item[Hadronic interaction model (relevant for the full energy spectrum).] The hadronic interaction model is a dominant source of systematic uncertainty in $\nu_{\rm atm}$ flux calculations across the full energy range. The current work employs a combination of JAM and modified-DPMJET-III models. Future updates can significantly improve accuracy by incorporating recent experimental data. In the low-energy region (sub-GeV), precise hadron production measurements from the NA61/SHINE experiment at CERN can be used to tune pion and kaon production cross-sections, directly constraining the neutrino yields~\cite{NA61SHINE:2013utd, NA61SHINE:2011tlp}. A recent study by Sato et al.~\cite{Sato:2026avb} has demonstrated the feasibility of this approach, achieving 7--9\% flux uncertainty below 1~GeV by using NA61/SHINE data to directly tune hadronic interaction models. In the high-energy region (multi-TeV), forward physics data from LHC experiments (e.g., LHCf)~\cite{LHCf:2017fnw} provide critical constraints on particle production in the very forward region, which is particularly relevant for air showers. We plan to integrate these latest datasets to refine the hadronic interaction models used in our simulation, thereby reducing the overall systematic uncertainty in the flux predictions.
\end{description}

The refinements outlined above will be prioritized in future work as the necessary data and models become available. Importantly, these updates are designed to reduce the systematic uncertainties in their respective energy regions. By incorporating these refinements, we aim to provide a more robust and precise calculation of the $\nu_{\rm atm}$ flux, particularly improving the predictions in the low-to-intermediate energy range ($E_\nu \lesssim 100$~GeV).

\section{Summary}\label{sec:summary}

We have presented a comprehensive 3D $\nu_{\rm atm}$ flux calculation based on the HKKMS15 framework, with several key improvements. For the first time, the propagation of $\mu_{\rm atm}$ inside the Earth and their subsequent decay or nuclear capture are incorporated, producing a supplementary flux of low-energy neutrinos below $\sim 100$~MeV. All essential input models are updated to reflect recent experimental measurements: the AMS02-based primary cosmic ray model, IGRF2020 for the geomagnetic field, and the muon-recalibrated hadronic interaction model. The calculation is performed for a global network of seven detector sites (JUNO, SK, CJPL, KM3NeT/ORCA, IceCube, DUNE, TRIDENT), covering an energy range from 10~MeV to $10^4$~GeV.

For $E_\nu > 100$~MeV, the predicted fluxes at the seven sites converge above $\sim 10$~GeV, while significant site-dependent deviations appear at lower energies---driven by the different geomagnetic cutoff rigidities. For example, the $\nu_\mu$ flux at IceCube and DUNE is approximately twice that at JUNO below 1~GeV. The zenith-angle and azimuth-angle dependence of the flux is governed by the production geometry of $\nu_{\rm atm}$: downward-going neutrinos originate from local production regions directly above the detector, horizontal neutrinos from regions near the detector with elongated atmospheric paths, and upward-going neutrinos from the antipodal side of the Earth. Compared with the HKKMS15 calculation, the current fluxes show deviations of 2--10\%, which are traced to the spectral change in the primary cosmic ray model (higher below $\sim 40$~GeV, lower above $\sim 40$~GeV) and the recalibration of the hadronic interaction model.

For $E_\nu < 100$~MeV, the muon propagation contribution is found to be a global effect: the absolute increase in flux is approximately site-independent, while the fractional enhancement varies inversely with the baseline flux at each site. The resulting flavor hierarchy follows $\nu_e \approx \bar{\nu}_\mu > \nu_\mu > \bar{\nu}_e$, driven by the charge asymmetry of $\mu_{\rm atm}$ ($\mu^+ > \mu^-$) that stop inside the Earth. The nuclear capture channel produces $\nu_\mu$ with an energy spectrum extending to $\sim 95$~MeV, well above the Michel endpoint at $52.8$~MeV, directly impacting the DSNB and dark matter search regions.

The hadronic uncertainty $\delta_{\rm had}$ has been re-estimated using the muon-constrained method of Honda et al., achieving a significant improvement across the energy range: with the high-altitude muon constraint, $\delta_{\rm had}$ is reduced to $\sim 3$--8\% for $\nu_\mu$/$\bar{\nu}_\mu$ and $\sim 4$--10\% for $\nu_e$/$\bar{\nu}_e$ below 1~GeV, and to $\sim 3$--5\% in the horizontal direction over 1--10~GeV. The total uncertainty $\delta_{\rm tot}$ is reduced to $\sim 5$--7\% in the region where $\delta_{\rm had}$ is minimized (1--10~GeV, horizontal), with further improvements below 1~GeV under the high-altitude scenario.

The precise flux predictions presented in this work, particularly the first-time results below 100~MeV, provide essential inputs for ongoing and upcoming neutrino and rare-event physics programs. For neutrino oscillation studies, the 3D flux predictions and reduced uncertainties can serve as flux inputs for any atmospheric neutrino experiment. The seven-site framework presented here directly supports JUNO~\cite{JUNO:2015zny}, Super-Kamiokande~\cite{Super-Kamiokande:2017yvm}, Hyper-Kamiokande~\cite{Hyper-Kamiokande:2018ofw}, DUNE~\cite{DUNE:2020ypp}, KM3NeT/ORCA~\cite{Aiello:2024}, IceCube~\cite{IceCube:2014flw}, and TRIDENT~\cite{Ye:2023dch}, with site-specific fluxes available for each location, while the comprehensive geomagnetic coverage enables cross-experiment calibration and interpolation to other sites. For DSNB searches, the precisely calculated atmospheric neutrino flux provides the dominant background for current and future experiments including JUNO, SK-Gd~\cite{Sekiya:2021}, and Hyper-Kamiokande. For direct dark matter detection, the precise low-energy flux predictions constrain the neutrino fog that limits the sensitivity of experiments such as LZ~\cite{LZ:2022lsv}, XENONnT~\cite{XENON:2023cxc}, and PandaX-4T~\cite{PandaX-4T:2021bab,OHare:2021utq}. For indirect dark matter searches---spanning high-energy neutrinos from WIMP annihilation at IceCube~\cite{IceCube:2016dgk} to low-energy neutrinos from light dark matter annihilation at JUNO~\cite{JUNO:2023vyz}, Borexino~\cite{Borexino:2019wln}, and SK~\cite{Super-Kamiokande:2020sgt}---the updated fluxes improve the background characterization. For nucleon decay searches, the better-constrained atmospheric neutrino background directly impacts the sensitivity of SK~\cite{Super-Kamiokande:2020wjk}, JUNO~\cite{JUNO:2022qgr}, Hyper-Kamiokande, and DUNE. Future refinements, including local mountain profiles, time-dependent solar modulation, and updated atmospheric and hadronic interaction models, will further improve the accuracy of these predictions.

\section*{Acknowledgements}
This paper is dedicated to the late Professor Morihiro Honda, who passed away in November 2021. Honda-san is a distinguished neutrino scientist focusing on the calculation of precision atmospheric neutrino fluxes. His generosity in sharing his code and his significant contributions to the early stages of flux calculations at the JUNO site were invaluable to this research. His contributions and enduring spirit will continue to inspire our work.
This work was partially supported by the National Natural Science Foundation of China (Grants No.~12405125, ~12125506), by CAS Project for Young Scientists in Basic Research (Grant No.~YSBR-099), by the National Key R\&D Program of China (Grant No.~2024YFE0110500), by the Fundamental Research Funds for the  Central Universities (2026MS077).


\section*{Atmospheric flux model data availability}
The data from the 3D atmospheric flux model for the different experimental sites are provided at: \url{https://github.com/JIECheng2021/atm_nu_flux_data}.
More flux model data for Lake Baikal, Gran Sasso, and SNO Lab is under preparation and will be provided at the same link.


\newpage
\appendix
\section{Detailed Description of Input Models and Muon Propagation}\label{app:models}

This appendix provides the complete description of the input models and the muon propagation treatment, moved from Sec.~\ref{sec:framework} to streamline the main text.

\subsection{Primary Cosmic Ray Model}
\label{app:crmodel}

\begin{figure}[!t]
\begin{center}
\includegraphics[width=0.8\textwidth]{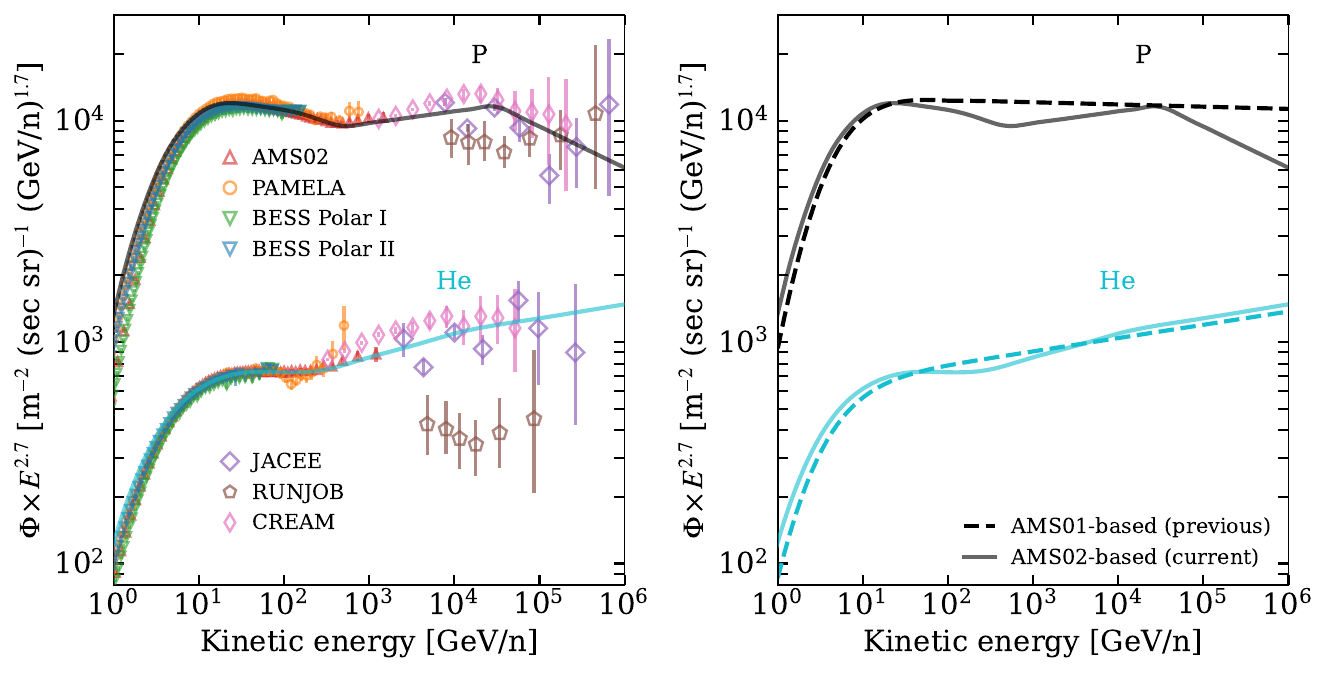}
\end{center}
\vspace{-0.8cm}
\caption{Primary cosmic ray data and spectra models. Left panel: Available data for cosmic ray proton and helium, along with the current AMS02-based primary cosmic ray model for solar minimum conditions (solid curves). Right panel: comparison between the current AMS02-based primary cosmic ray model for solar minimum conditions (solid curves) and the previous AMS01-based model (dashed curves).}
\label{fig:crmodel}
\end{figure}

For the HKKMS11 and HKKMS15 calculations, the primary flux model is based on AMS01~\cite{AMS:2000mzk,AMS:2000hdi} and BESS~\cite{Haino:2004nq,Sanuki:2000wh} observations, with a spectral index of $-2.71$ above 100~GeV~\cite{Gaisser:2002jj}, shown as dashed lines in the right panel of Fig.~\ref{fig:crmodel}, where the AMS02-based solar-minimum model is also shown for comparison (solid lines). Introduced in Ref.~\cite{Honda:2004yz}, this model is referred to as the ``AMS01-based primary cosmic ray model''.

As illustrated in the left panel of Fig.~\ref{fig:crmodel}, recent data from AMS02~\cite{AMS:2015tnn,AMS:2015azc}, BESS-polar~\cite{Abe:2015mga} and PAMELA~\cite{PAMELA:2011mvy} have become available, showing substantial agreement, especially from AMS02, with only a few percent uncertainty below 100~GeV. The experimental data are distinguished by different symbols as labeled in the figure legend. Above a few TeV, JACEE~\cite{Christ:1998zz}, RUNJOB~\cite{RUNJOB:2005mtb} and CREAM~\cite{Yoon:2017qjx} data are available, but they exhibit significant scatter among the experimental groups.

The new AMS02-based primary cosmic ray model, depicted as solid lines in Fig.~\ref{fig:crmodel}, is primarily based on AMS02 data below 1~TeV, which boasts the highest statistics in this energy region and shows excellent agreement with the BESS-polar data below 100~GeV once both datasets are corrected to solar minimum conditions. Additionally, PAMELA agrees with AMS02 within 5\% below 200~GeV, with exceptions noted below 10~GeV due to solar modulation effects.

Above 1~TeV, the flux values for proton cosmic rays from JACEE, RUNJOB, and CREAM are consistent at 10~TeV, falling within each dataset's error bars. Therefore, these data are used to construct the spectrum model for proton cosmic rays. In contrast, helium flux data from JACEE, RUNJOB, and CREAM exhibit greater variability than the proton data. Consequently, only the CREAM and JACEE data in this energy range are utilized to develop the helium cosmic ray model. Note that other heavy nuclei in primary cosmic rays are also under consideration and will be analyzed using recent measurements.

The right panel of Fig.~\ref{fig:crmodel} compares the AMS02-based primary cosmic ray model to the AMS01-based model. Below 40~GeV, the AMS02-based model shows higher cosmic ray flux values, with the increase tapering off as energy rises. Above 40~GeV, the current model exhibits lower flux values for primary protons, with a maximum discrepancy of $\sim 50\%$. These changes affect the calculated $\nu_{\rm atm}$ flux in a straightforward manner: the increased low-energy primary flux leads to a higher $\nu_{\rm atm}$ flux below a few GeV, while the reduced high-energy primary flux results in a lower $\nu_{\rm atm}$ flux at higher energies. A detailed comparison of the $\nu_{\rm atm}$ flux variations between the two primary cosmic ray models is presented in Sec.~\ref{subsec:comparison_hkkms15}.

In the 3D calculation framework, the primary cosmic ray model serves as the input flux spectrum ($\Phi_{\mathrm{cr}}$) for particles sampled on the injection sphere, directly determining the overall normalization and energy dependence of the resulting $\nu_{\rm atm}$ flux.

\subsection{Geomagnetic Field Model}
\label{app:geomagnetic}

\begin{figure}[!t]
\begin{center}
\includegraphics[width=0.9\textwidth]{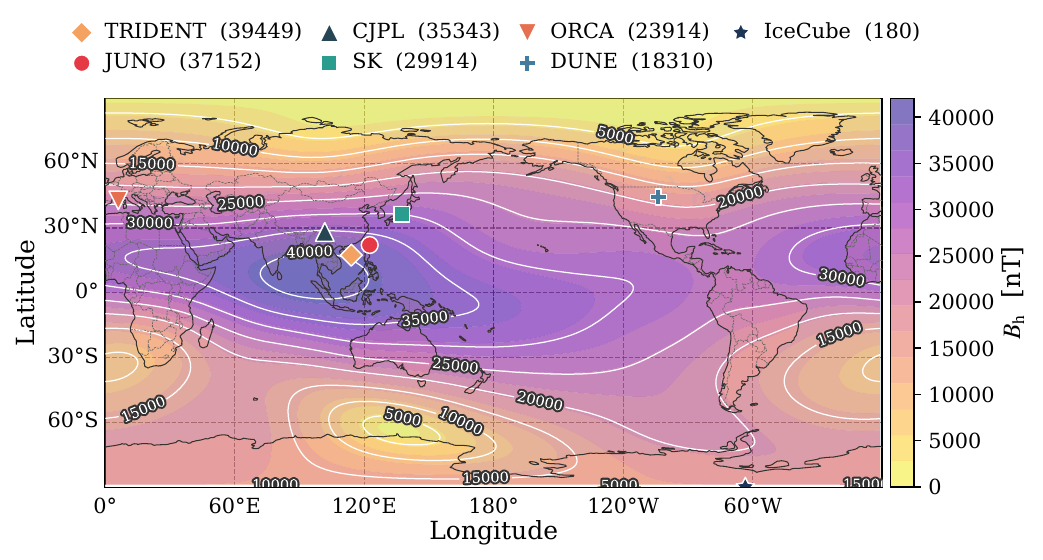}
\end{center}
\vspace{-0.8cm}
\caption{The geomagnetic horizontal field strength obtained from IGRF2020, along with data from the sites JUNO, SK, CJPL, KM3NeT/ORCA, IceCube, DUNE and TRIDENT, where the $\nu_{\rm atm}$ flux is calculated. }
\label{fig:geomagnetic_field}
\end{figure}
The International Geomagnetic Reference Field (IGRF) model~\cite{IGRF13} is utilized in the 3D $\nu_{\rm atm}$ flux calculation. This model consists of a set of spherical harmonic coefficients that are integrated into a mathematical framework to describe the large-scale, time-varying component of Earth's internal magnetic field. Updated approximately every five years based on new observational data, the model has evolved through various versions, including IGRF2005 and IGRF2010, which were used in HKKMS11 and HKKMS15, respectively. For the current work, we have employed IGRF2020.

The geomagnetic field affects the $\nu_{\rm atm}$ flux through two distinct mechanisms: the \textit{rigidity cutoff}, which filters incoming primary cosmic rays before they reach the atmosphere, and \textit{muon bending}, which deflects the trajectories of charged secondary particles produced in atmospheric cascades. Both effects must be accurately modeled to obtain reliable $\nu_{\rm atm}$ flux predictions at a given geographic location.

To understand the rigidity cutoff, consider the cyclotron motion of a primary cosmic ray with momentum $p$ and charge $Ze$ traversing the geomagnetic field $B$. The gyroradius is $R_g = p/(ZeB)$. If $R_g$ is smaller than the Earth's radius, the particle cannot penetrate to low altitudes and is effectively excluded from the atmosphere. As $R_g$ increases, the cosmic ray can access closer points, but the Earth acts as a tilted boundary that restricts its access azimuth angle. When $R_g$ becomes sufficiently large, the Earth behaves as an effectively flat boundary that blocks only upward-going cosmic rays, and the azimuthal restriction vanishes. This defines the rigidity cutoff at a given location and arrival direction.

The muon bending effect arises because the geomagnetic field deflects charged muons as they propagate through the atmosphere. Since the $\nu_{\rm atm}$ flux above a few GeV depends strongly on the arrival zenith angle, even a small angular deviation caused by muon bending can lead to a substantial change in the calculated flux. The horizontal component of the geomagnetic field ($B_h$) is the primary contributor to this effect: the vertical component produces a Lorentz force parallel to the particle's transverse motion and does not alter the arrival direction, whereas $B_h$ generates a transverse force that shifts the muon trajectory perpendicular to its original path, thereby changing the apparent arrival zenith angle at the detector.

In the 3D calculation framework, the IGRF2020 model is incorporated into the backward trajectory tracing of each sampled cosmic ray. The equation of motion is solved backward in time through the geomagnetic field described by IGRF2020, and a particle is accepted if its backward trajectory reaches the outer simulation boundary without re-crossing the injection sphere. This procedure naturally accounts for both the rigidity cutoff and the directional deflections induced by the geomagnetic field.

Figure~\ref{fig:geomagnetic_field} illustrates the horizontal component of the geomagnetic field based on the IGRF2020 model, highlighting the locations of the seven detector sites where the $\nu_{\rm atm}$ flux is calculated: JUNO ($B_h \sim 37\,000$~nT), SK ($B_h \sim 30\,000$~nT), CJPL ($B_h \sim 35\,000$~nT), TRIDENT ($B_h \sim 39\,000$~nT), KM3NeT/ORCA ($B_h \sim 24\,000$~nT), DUNE ($B_h \sim 18\,000$~nT), and IceCube ($B_h \sim 200$~nT). The wide range of field strengths --- from near-zero at the South Pole (IceCube) to $\sim\!40\,000$~nT at low-latitude sites (TRIDENT, JUNO) --- results in markedly different rigidity cutoffs and muon bending effects across the sites, underscoring the importance of an accurate geomagnetic field model in multi-site $\nu_{\rm atm}$ flux calculations.

\subsection{Atmospheric Model}
\label{app:atmospheric}

The US-standard '76 atmospheric model~\cite{US76} has been widely used in $\nu_{\rm atm}$ flux studies, including in HKKMS11 and earlier calculations. However, this model only represents air density as a function of altitude, lacking time variation and position dependence across the Earth. Since HKKMS15, the NRLMSISE-00 global atmospheric model~\cite{NR00}, which accounts for position dependence and temporal variations, has been utilized to calculate atmospheric neutrino flux.

In the 3D calculation framework, the atmospheric model determines the air density profile along the trajectories of primary cosmic rays and their secondary particles. The air density at each altitude directly affects the competition between meson decay and re-interaction: at higher densities, pions and kaons are more likely to undergo hadronic re-interactions rather than decay, suppressing the resulting $\nu_{\rm atm}$ flux. This effect is particularly important for low-energy mesons ($E \lesssim 10$~GeV), whose decay lengths are comparable to the atmospheric scale height, and for forward-going particles that traverse larger atmospheric columns.

In Ref.~\cite{Honda:2015fha}, the $\nu_{\rm atm}$ flux calculated using the NRLMSISE-00 atmospheric model is compared to that obtained with the US-standard '76 atmospheric model at various locations, including the India-based Neutrino Observatory (INO) site for tropical and equatorial regions, as well as the South Pole and Pyh\"asalmi mine (Finland) for polar regions, and the SK site. The NRLMSISE-00 atmospheric model closely aligns with the US-standard '76 model in the tropical and mid-latitude regions, but diverges significantly in the polar region. Additionally, the NRLMSISE-00 model indicates substantial seasonal variation in the polar areas. The
comparison at the SK site demonstrates that the $\nu_{\rm atm}$ flux difference between the two atmospheric models is small (a few percent) in
mid-latitude regions, consistent with the similarity of their air density profiles at these locations.

In the current calculation, we continue to use the NRLMSISE-00 model for its ability to provide site- and time-dependent atmospheric descriptions across the wide range of geographic locations covered in this work. These include subtropical sites (JUNO, TRIDENT), mid-latitude sites (CJPL, SK, KM3NeT/ORCA, DUNE), and the polar site IceCube at the South Pole. For subtropical and mid-latitude sites, the seasonal variation in the atmospheric density profile is modest, and the resulting systematic uncertainty in the $\nu_{\rm atm}$ flux from the atmospheric model is estimated to be $\sim 3\%$, based on a $\pm 5\%$ variation in the air density profile to account for seasonal
changes~\cite{Honda:2006qj}. For the polar site (IceCube), NRLMSISE-00 captures the substantial seasonal variation characteristic of high-latitude regions, which the US-standard '76 model cannot describe.

\subsection{Hadronic Interaction Model}
\label{app:hadronic}

For the hadronic interaction model, theoretically constructed models are employed in the 3D $\nu_{\rm atm}$ flux calculation, which have proven effective in detector simulations for high-energy accelerator experiments. In the 3D calculation framework, the hadronic interaction model determines the neutrino and muon yields ($Y_{\nu}$ and $Y_{\mu}$) by simulating the production and propagation of secondary particles (primarily pions and kaons) in atmospheric cascades. Given that $\mu_{\rm atm}$'s are primarily produced through pion decay processes, their measurements serve as crucial calibrated sources for validating the hadronic interaction models. The accurately measured $\mu_{\rm atm}$ fluxes from various sources --- such as the BESS detector at Tsukuba (sea level)~\cite{Haino:2004nq}, at
Mt.~Norikura (2770~m~a.s.l)~\cite{Sanuki:2002tp}, the L3+C experiment at CERN~\cite{L3:2004sed}, and the horizontal muon flux data from the MUTRON experiment~\cite{Matsuno:1984kq} --- have been used to constrain the uncertainties in the $\nu_{\rm atm}$ flux calculation.

Since HKKMS11, the combination of modified DPMJET-III~\cite{Honda:2006qj} for energies above 32~GeV and JAM, a nuclear interaction model developed with PHITS (Particle and Heavy-Ion Transport code System)~\cite{Niita:2006zz}, for energies below 32~GeV has been used to simulate hadronic interactions. This combination offers a better reproduction of the observed muon spectra compared to other combinations~\cite{Honda:2011nf}.

The muon calibration has been repeated following the updated AMS02-based primary cosmic ray flux model. Using the above accurate muon observations as a reference, the DPMJET-III and JAM interaction models have been modified in accordance with the AMS02-based primary cosmic ray model. The resulting modified interaction model is referred to as the ``muon-recalibrated interaction model''~\cite{Honda:2019ymh}. With the previous interaction model (used in HKKMS11 and HKKMS15), the discrepancy between calculated and observed $\mu_{\rm atm}$ fluxes exceeded 10\% around 100~GeV when combined with the AMS02-based primary cosmic ray model. After applying the muon-recalibrated interaction model, this discrepancy was reduced to within 5\% over the 1--100~GeV range. The muon-recalibrated hadronic interaction model has been implemented in this work.

For $\nu_{\rm atm}$ with energies above a few GeV, the hadronic interaction model is constrained by the observed $\mu_{\rm atm}$ flux, assuming a similarity in the meson production density distribution for both $\nu_{\rm atm}$ and $\mu_{\rm atm}$ within the phase space of hadronic interactions. However, this assumption is not valid for $\nu_{\rm atm}$ below 1~GeV due to the energy loss of muons in the atmosphere, which breaks the direct link between the observed surface muon flux and the neutrino production phase space. The systematic uncertainty in this low-energy region is discussed further in Sec.~\ref{sec:uncertainty}.

\subsection{Muon Propagation Inside the Earth}
\label{app:muon}

\begin{figure}[!t]
\begin{center}
\includegraphics[width=0.6\textwidth]{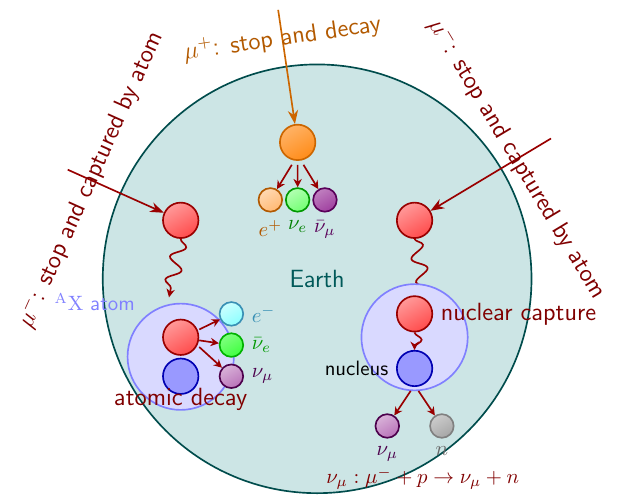}
\end{center}
\vspace{-0.8cm}
\caption{Schematic diagram of muon decay and capture processes inside the Earth. A stopped $\mu^{+}$ decays freely into $e^{+}\nu_{e}\bar{\nu}_{\mu}$ (left). A stopped $\mu^{-}$ is captured into a muonic atom, where it either decays in orbit ($e^{-}\bar{\nu}_{e}\nu_{\mu}$) or undergoes nuclear capture (emitting $\nu_{\mu}$ only), with the probability determined by the element-dependent branching fractions listed in Tab.~\ref{table:tab2}.}
\label{fig:muinEarth}
\end{figure}

For $\nu_{\rm atm}$ flux below 100~MeV, neutrinos generated by muon propagation inside the Earth make a significant contribution. However, accounting for muon decay inside the Earth poses challenges and is absent in the HKKMS11 and HKKMS15 calculations. When muons enter the Earth, they rapidly lose energy through electromagnetic interactions and eventually come to a stop. For most $\mu_{\rm atm}$, their travel distance inside Earth is less than 100~m. When muons come to a stop, various decay or capture processes may occur within the Earth, as illustrated in Fig.~\ref{fig:muinEarth}. A $\mu^{+}$ decays freely into an $e^{+}$, a $\nu_{e}$ and a $\bar{\nu}_{\mu}$, and this process is well understood. In contrast, a $\mu^{-}$ is rapidly captured by an atom, forming a muonic atom. Once bound, the $\mu^{-}$ has two possible fates: it can either decay in orbit into an $e^{-}$, a $\bar{\nu}_{e}$ and a $\nu_{\mu}$, or it can be captured by a bound proton in the nucleus, emitting a $\nu_{\mu}$. The competition between these two channels is governed by the element-dependent decay probability $D_{\mu^{-}}$ (and nuclear capture probability $1-D_{\mu^{-}}$), listed in Tab.~\ref{table:tab2}. For water-dominant elements such as oxygen, $D_{\mu^{-}} \approx 81.6\%$, meaning most $\mu^{-}$ decay in orbit; for rock-forming elements such as silicon, $D_{\mu^{-}} \approx 34.1\%$, so nuclear capture dominates. At present, the nuclear capture of $\mu^{-}$ remains one of the least understood aspects of muon decay in matter.

\renewcommand{\arraystretch}{1.2}

\begin{table}[!tb]
\centering
\caption{
The probabilities of $\mu^{-}$ atomic capture and decay in muonic atoms ($D_{\mu^{-}}$), as well as the mean lifetime of $\mu^{-}$ ($\tau_{\mu^{-}}$), are listed for the ten most abundant elements of the upper continental crust, along with their respective number percentages. These values are taken from Ref.~\cite{Guo:2018sno}.}
\begin{tabular}{lcccc}
\hline \hline
Elements &  Number (\%)  &  Atomic capture (\%)  & $\tau_{\mu^{-}}$ (ns)  & $D_{\mu^{-}}$ (\%)\\
\hline
O &62.13 & 60.26 & 1795.4 & 81.56 \\
Si & 23.89 & 19.46 & 756 & 34.14 \\
Al &3.91 & 2.88 & 864 & 39.05 \\
Fe &2.27 & 7.21 &206 & 9.14 \\
Ca &2.07 &3.81 &332.7 &14.92 \\
Na &2.27 &2.21 &1204  &54.58 \\
K  &1.28 &1.91 &435   &19.54 \\
Mg &1.99 &1.79 &1067.2&48.33 \\
Ti &0.17 &0.45 &329.3 &14.70 \\
P  &0.02 &0.02 &611.2 &27.57 \\
\hline \hline
\end{tabular}
\label{table:tab2}
\end{table}

\begin{figure}[!t]
\begin{center}
\includegraphics[width=0.9\textwidth]{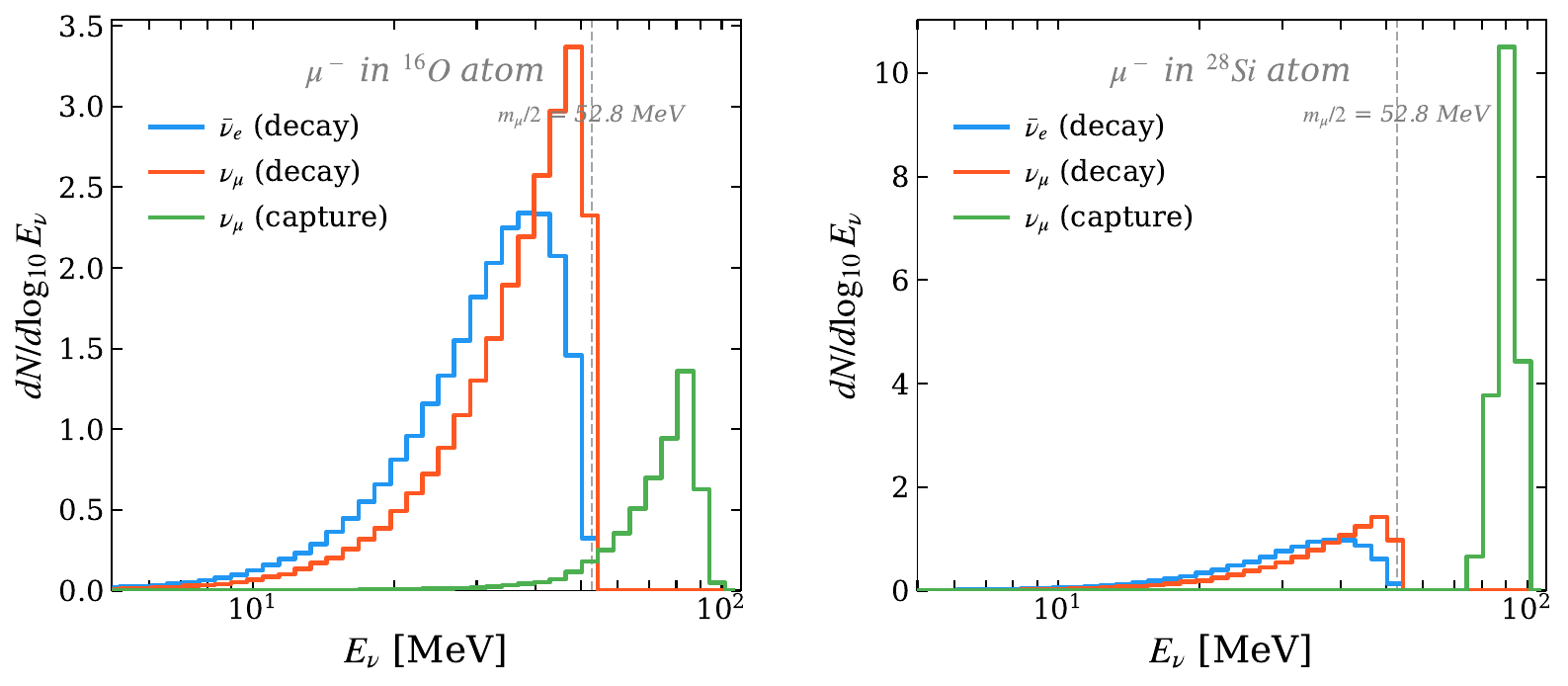}
\end{center}
\vspace{-0.8cm}
\caption{Energy spectra ($dN/d\log_{10}E$) of neutrinos from $\mu^{-}$ absorbed in $^{16}$O atom (left panel) and $^{28}$Si atom (right panel). Three components are shown: $\bar{\nu}_e$ from muon decay (blue), $\nu_\mu$ from muon decay (red), and $\nu_\mu$ from nuclear capture (green). The gray dashed line marks the Michel endpoint $m_\mu/2 \approx 52.8$~MeV, which limits the decay spectra; in contrast, the nuclear capture spectrum extends to $\sim$95~MeV. The integral of each curve equals the corresponding neutrino yield per absorbed $\mu^{-}$, weighted by the branching fractions (O: 81.6\% decay / 18.4\% capture; Si: 34.1\% decay / 65.9\% capture).}
\label{fig:mucapspe}
\end{figure}

\begin{figure}[!t]
\begin{center}
\includegraphics[width=0.9\textwidth]{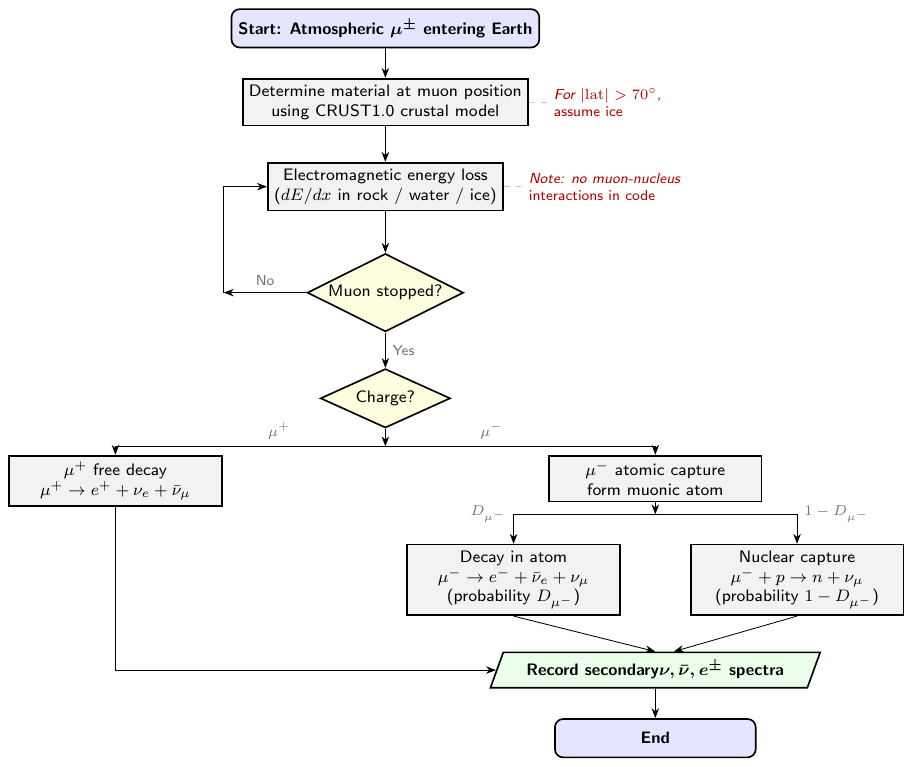}
\end{center}
\vspace{-0.8cm}
\caption{Flowchart of the muon propagation calculation. The surrounding medium (rock, water, or ice) is determined from the CRUST1.0 global crustal model. Each muon is propagated via electromagnetic energy loss until it stops. The charge sign then determines the pathway: $\mu^{+}$ decays freely, while $\mu^{-}$ is captured into a muonic atom where it either decays in orbit or undergoes nuclear capture. The resulting particles ($\nu_{e}, \bar{\nu}_{e}, \nu_{\mu}, e^{\pm}$) are shown at each terminal.}
\label{fig:muworkflow}
\end{figure}

\begin{figure}[!t]
\begin{center}
\includegraphics[width=0.8\textwidth]{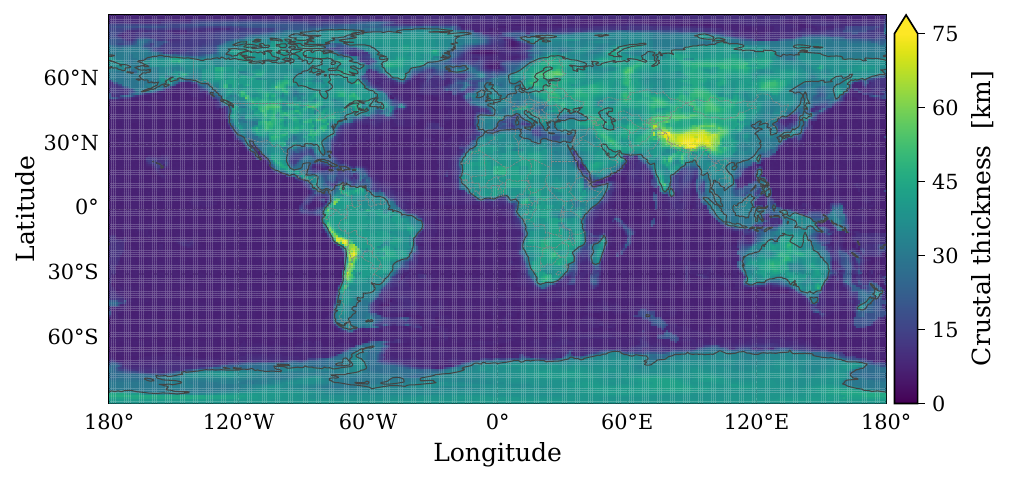}
\end{center}
\vspace{-0.8cm}
\caption{Global map of crustal thickness based on the CRUST1.0 model~\cite{crust}. The map distinguishes thick continental crust (30--70~km, e.g., under China, Europe, North America) from thin oceanic crust (5--10~km), which determines whether a stopping muon propagates through rock or water. For latitudes $> 70^{\circ}$ and $< -70^{\circ}$, the surface is assumed to be covered by ice.}
\label{fig:curst1}
\end{figure}

We have conducted a general study of muon decay in Earth's material, building upon the work in Ref.~\cite{Guo:2018sno}. We treat the upper layer of the Earth as composed of rock and water to study the decay of $\mu^{-}$ in both environments. For $\mu^{-}$ decays in the rock, Ref.~\cite{Guo:2018sno} provides the probabilities of $\mu^{-}$ atomic capture and decay in muonic atoms for the ten most abundant elements of the upper continental crust, as presented in Tab.~\ref{table:tab2}. The probabilities of $\mu^{-}$ atomic capture for these ten elements are approximately proportional to their respective number percentages. In muonic atoms, $\mu^{-}$ decay and nuclear capture compete with each other. Assuming the lifetime of $\mu^{-}$ decay in muonic atoms is similar to that of free $\mu^{-}$ decay ($\sim$ 2.2~$\mu$s), the decay probability $D_{\mu^{-}}$ and the nuclear capture probability ($1-D_{\mu^{-}}$) can be estimated given the known mean lifetime of $\mu^{-}$ for these elements. For $\mu^{-}$ decays in the water, we consider only oxygen element, assuming an atomic absorption probability of 100\%. 

For a stopped muon that decays---whether $\mu^{+}$ freely or $\mu^{-}$ in a muonic atom---the normalized ${\nu}_{e}/\bar{\nu}_{e}$ and ${\nu}_{\mu}/\bar{\nu}_{\mu}$ energy spectra follow the well-known Michel spectrum and can be written as 
\begin{equation}
    f_{{\nu}_{e}/\bar{\nu}_{e}} = \frac{192}{m_{\mu}}\left[\left(\frac{E_{\nu}}{m_{\mu}}\right)^{2}\left(\frac{1}{2}-\frac{E_{\nu}}{m_{\mu}}\right)\right], 
    \label{Eq:eq2}
\end{equation}
\begin{equation}
    f_{{\nu}_{\mu}/\bar{\nu}_{\mu}} = \frac{64}{m_{\mu}}\left[\left(\frac{E_{\nu}}{m_{\mu}}\right)^{2}\left(\frac{3}{4}-\frac{E_{\nu}}{m_{\mu}}\right)\right], 
    \label{Eq:eq3}
\end{equation}
where $m_{\mu}$ is the muon mass and $E_{\nu} \le m_{\mu}/2$. For the nuclear capture of $\mu^{-}$, there are currently no theoretical or experimental energy spectra for $\nu_{\mu}$. However, the energy spectra of $\nu_{\mu}$ closely resemble those of the $\gamma$ spectra from the capture of $\pi^{-}$ by nuclei. As suggested in Ref.~\cite{Measday:2001yr}, the $\gamma$ spectra from the reactions $^{16}$O$(\pi^{-},\gamma)^{16}$N$^{*}$~\cite{Strassner:1979pz} and $^{28}$Si$(\pi^{-},\gamma)^{28}$Al$^{*}$~\cite{Measday:2001yr} are used as proxies for the $\nu_{\mu}$ energy spectra from the nuclei capture. To derive the neutrino energy spectra, it is important to note that the maximal $\nu_{\mu}$ energy from nuclear capture is limited to $\sim$95~MeV, reflecting the available energy from the reaction $\mu^{-} + p \rightarrow \nu_{\mu} + n$ corrected for nuclear binding effects, in contrast to the Michel endpoint at $m_{\mu}/2 \approx 52.8$~MeV for in-atom muon decay. Figure~\ref{fig:mucapspe} presents the energy spectra ($dN/d\log_{10}E$) of neutrinos from $\mu^{-}$ absorption in $^{16}$O and $^{28}$Si, respectively. The three components are shown separately: $\bar{\nu}_e$ from in-atom muon decay (blue), $\nu_\mu$ from in-atom muon decay (red), and $\nu_\mu$ from nuclear capture (green). The gray dashed line marks the Michel endpoint $m_\mu/2 \approx 52.8$~MeV, and the branching fractions (O: 81.6\% decay / 18.4\% capture; Si: 34.1\% decay / 65.9\% capture) are annotated directly on the figure. We observe a significant difference in the $\nu_{\mu}$ capture spectra between O and Si, primarily attributed to their different nuclear-capture probabilities. In addition, there are no other experimental data for $\pi^{-}$ captured on heavier nuclei above Si. Therefore, we have used the results of Si as a proxy for these heavier nuclei. However, given the differences in the $\gamma$-spectra from $\pi^{-}$ absorption in O and Si, the discrepancies for heavier nuclei compared to Si may be significant. Further studies will be necessary when relevant data become available.

The calculation procedure for muon propagation in the Earth is summarized in the flowchart of Fig.~\ref{fig:muworkflow}. For each muon entering the Earth, the surrounding medium (rock, water, or ice) is first determined from the CRUST1.0 global crustal model~\cite{crust}. The muon is then propagated through electromagnetic energy loss until it stops. Once stopped, the charge sign determines the pathway: $\mu^{+}$ decays freely, while $\mu^{-}$ is captured into a muonic atom. In the muonic atom, the $\mu^{-}$ either decays in orbit (with element-dependent probability $D_{\mu^{-}}$) or undergoes nuclear capture (with probability $1-D_{\mu^{-}}$), producing the corresponding neutrinos. Note that this calculation does not account for muon-nucleus interactions during propagation; we assume all muons stop purely through electromagnetic energy loss. In the global simulation, we have used crustal thickness data from the CRUST1.0 model, as shown in Fig.~\ref{fig:curst1}, which maps the global distribution of thick continental crust (30--70~km) and thin oceanic crust (5--10~km). This distinction is critical for identifying whether the medium along a muon trajectory is rock or water, which directly affects the energy-loss rate and the $\mu^{-}$ nuclear capture probability.

\end{document}